\documentclass[%
superscriptaddress,
twocolumn,
showpacs,
nofootinbib,
 amsmath,amssymb,
 aps,
 prd,
floatfix,
]{revtex4-1}

\usepackage{amsmath}
\usepackage{float}
\usepackage{empheq}
\usepackage{xcolor}
\usepackage{graphicx}
\usepackage{dcolumn}
\usepackage{bm}
\usepackage{hhline} 
\usepackage{siunitx}

\usepackage{multirow}
\usepackage[normalem]{ulem}

\usepackage{hyperref}
\hypersetup{
    pdfnewwindow=true,      
    colorlinks=true,       
    linkcolor=blue,          
    citecolor=blue,        
    filecolor=blue,      
    urlcolor=blue           
}

\usepackage[utf8]{inputenc}
\usepackage{longtable}
\usepackage[inline]{enumitem}
\usepackage{array}

\renewcommand{\vec}[1]{\mathbf{#1}}
\let\oldhat\hat
\renewcommand{\hat}[1]{\oldhat{\mathbf{#1}}}

\newcommand{\enu}{E_{\nu}}
\newcommand{\difd}{\textrm{d}}
\newcommand{\dEdx}{\difd E/\difd x}
\newcommand{\proton}{\textrm{p}}
\newcommand{\pproton}{p_\proton}
\newcommand{\thetaproton}{\theta_\proton}
\newcommand{\pmu}{p_\mu}
\newcommand{\thetamu}{\theta_\mu}
\newcommand{\neutron}{\textrm{n}}

\newcommand{\pn}{p_\neutron}

\newcommand{\dalphat}{\delta\alpha_\textrm{T}}
\newcommand{\pt}{p_\textrm{T}}
\newcommand{\dpt}{\delta \pt}
\newcommand{\dphit}{\delta \phi_\textrm{T}}

\newcommand{\dptx}{\delta p_\textrm{Tx}}
\newcommand{\dpty}{\delta p_\textrm{Ty}}
\newcommand{\dptyavg}{\langle\dpty\rangle}
\newcommand{\dptv}{\delta\vec{p}_\textrm{T}}
\newcommand{\dptvh}{\hat{\delta\vec{p}}_\textrm{T}}

\newcommand{\gev}{~\textrm{GeV}}
\newcommand{\gevc}{~\textrm{GeV}/c}
\newcommand{\gevcc}{~\textrm{GeV}/c^2}
\newcommand{\mev}{~\textrm{MeV}}
\newcommand{\mevc}{~\textrm{MeV}/c}

\newcommand{\ptl}{\vec{p}_\textrm{T}^{\,\mu}}
\newcommand{\ptlh}{\hat{\vec{p}}_\textrm{T}^{\,\mu}}
\newcommand{\pnuh}{\hat{\vec{p}}_{\nu}}
\newcommand{\ptlm}{p_\textrm{T}^{\,\mu}}

\newcommand{\ptni}{\vec{p}_\textrm{T}^{\,\proton}}

\newcommand{\genie}{\textsc{genie}}
\newcommand{\gver}{2.8.4}
\newcommand{\gvernew}{2.12.10}
\newcommand{\nuwro}{\textsc{n}u\textsc{w}ro}
\newcommand{\neut}{\textsc{neut}}
\newcommand{\gibuu}{\textsc{g}i\textsc{buu}}
\newcommand{\nuisance}{\textsc{nuisance}}

\newcommand{\mnvgenie}[1]{\textsc{m}nv\genie-v#1}

\newcommand{\ufsi}{U_\mathrm{opt}}
\newcommand{\veff}{V_\mathrm{eff}}
\newcommand{\veffp}{V_\mathrm{eff}^\mathrm{P}}
\newcommand{\Ma}{M_\mathrm{A}}
\newcommand{\ma}{M_\mathrm{A-1}}
\newcommand{\mas}{{M^*_\mathrm{A-1}}}
\newcommand{\ex}{E_x}
\newcommand{\exN}{{E_x^N}}

\newcommand{\dN}[1]{\Delta_\genie^\text{#1}}

\newcommand{\stki}{single-TKI}

\newcommand{\minerva}{$\textsc{MINER}\nu\textsc{A}$}
\newcommand{\minos}{$\textsc{MINOS}$}
\newcommand{\ttk}{$\textsc{T2K}$}
\newcommand{\microboone}{$\textsc{MicroBooNE}$}
\newcommand{\dune}{$\textsc{dune}$}

\newcommand{\atx}{A_\mathrm{Tx}}
\newcommand{\alr}{A_\mathrm{LR}}

\newcommand{\intE}{interaction energy}
\newcommand{\removalE}{removal energy}

\newcommand{\qzero}{{q_0}}
\newcommand{\mnprime}{{M_N^\prime}}

\newcommand{\range}[2]{(#1,#2)}
\begin{document}

\preprint{APS/123-QED}

\title{Nucleon binding energy and transverse momentum imbalance in neutrino-nucleus reactions}

\newcommand{\Rutgers}{Rutgers, The State University of New Jersey, Piscataway, New Jersey 08854, USA}
\newcommand{\Hampton}{Hampton University, Dept. of Physics, Hampton, VA 23668, USA}
\newcommand{\Dortmund}{Institute of Physics, Dortmund University, 44221, Germany }
\newcommand{\Otterbein}{Department of Physics, Otterbein University, 1 South Grove Street, Westerville, OH, 43081 USA}
\newcommand{\JMU}{James Madison University, Harrisonburg, Virginia 22807, USA}
\newcommand{\Florida}{University of Florida, Department of Physics, Gainesville, FL 32611}
\newcommand{\UCIrvine}{Department of Physics and Astronomy, University of California, Irvine, Irvine, California 92697-4575, USA}
\newcommand{\CBPF}{Centro Brasileiro de Pesquisas F\'{i}sicas, Rua Dr. Xavier Sigaud 150, Urca, Rio de Janeiro, Rio de Janeiro, 22290-180, Brazil}
\newcommand{\PUCP}{Secci\'{o}n F\'{i}sica, Departamento de Ciencias, Pontificia Universidad Cat\'{o}lica del Per\'{u}, Apartado 1761, Lima, Per\'{u}}
\newcommand{\INRM}{Institute for Nuclear Research of the Russian Academy of Sciences, 117312 Moscow, Russia}
\newcommand{\Jlab}{Jefferson Lab, 12000 Jefferson Avenue, Newport News, VA 23606, USA}
\newcommand{\Pittsburgh}{Department of Physics and Astronomy, University of Pittsburgh, Pittsburgh, Pennsylvania 15260, USA}
\newcommand{\Guanajuato}{Campus Le\'{o}n y Campus Guanajuato, Universidad de Guanajuato, Lascurain de Retana No. 5, Colonia Centro, Guanajuato 36000, Guanajuato M\'{e}xico.}
\newcommand{\Athens}{Department of Physics, University of Athens, GR-15771 Athens, Greece}
\newcommand{\Tufts}{Physics Department, Tufts University, Medford, Massachusetts 02155, USA}
\newcommand{\WM}{Department of Physics, College of William \& Mary, Williamsburg, Virginia 23187, USA}
\newcommand{\FNAL}{Fermi National Accelerator Laboratory, Batavia, Illinois 60510, USA}
\newcommand{\Purdue}{Department of Chemistry and Physics, Purdue University Calumet, Hammond, Indiana 46323, USA}
\newcommand{\MCLA}{Massachusetts College of Liberal Arts, 375 Church Street, North Adams, MA 01247}
\newcommand{\UMD}{Department of Physics, University of Minnesota -- Duluth, Duluth, Minnesota 55812, USA}
\newcommand{\Northwestern}{Northwestern University, Evanston, Illinois 60208}
\newcommand{\UNI}{Universidad Nacional de Ingenier\'{i}a, Apartado 31139, Lima, Per\'{u}}
\newcommand{\Rochester}{University of Rochester, Rochester, New York 14627 USA}
\newcommand{\Austin}{Department of Physics, University of Texas, 1 University Station, Austin, Texas 78712, USA}
\newcommand{\USM}{Departamento de F\'{i}sica, Universidad T\'{e}cnica Federico Santa Mar\'{i}a, Avenida Espa\~{n}a 1680 Casilla 110-V, Valpara\'{i}so, Chile}
\newcommand{\Geneva}{University of Geneva, 1211 Geneva 4, Switzerland}
\newcommand{\Chicago}{Enrico Fermi Institute, University of Chicago, Chicago, IL 60637 USA}
\newcommand{\hired}{}
\newcommand{\OregonState}{Department of Physics, Oregon State University, Corvallis, Oregon 97331, USA}
\newcommand{\oxford}{Oxford University, Department of Physics, Oxford, United Kingdom }
\newcommand{\umiss}{University of Mississippi, Oxford, Mississippi 38677, USA}
\newcommand{\upenn}{Department of Physics and Astronomy, University of Pennsylvania, Philadelphia, PA 19104}
\newcommand{\AMU}{AMU Campus, Aligarh, Uttar Pradesh 202001, India}
\newcommand{\wroclaw}{University of Wroclaw, plac Uniwersytecki 1, 50-137 Wrocław, Poland}
\newcommand{\Mohali}{IISER, Mohali, Knowledge city, Sector 81, Manauli PO 140306, Punjab, India}
\newcommand{\CINVESTAV}{Departamento de Fisica Col. San Pedro Zacatenco, 07360 Mexico, DF, Av. Instituto Politécnico Nacional, Mexico}
\newcommand{\york}{York University, Department of Physics and Astronomy, Toronto, Ontario, M3J 1P3 Canada}

\author{T.~Cai}                           \affiliation{\Rochester}
\author{X.-G.~Lu}                         \affiliation{\oxford}
\author{L.A.~Harewood}                    \affiliation{\UMD}
\author{C.~Wret}                          \affiliation{\Rochester}
\author{F.~Akbar}                         \affiliation{\AMU}
\author{D.A.~Andrade}                     \affiliation{\Guanajuato}
\author{M.~V.~Ascencio}                   \affiliation{\PUCP}
\author{L.~Bellantoni}                    \affiliation{\FNAL}
\author{A.~Bercellie}                     \affiliation{\Rochester}
\author{M.~Betancourt}                    \affiliation{\FNAL}
\author{A.~Bodek}                         \affiliation{\Rochester}
\author{J.~L.~Bonilla}                    \affiliation{\Guanajuato}
\author{A.~Bravar}                        \affiliation{\Geneva}
\author{H.~Budd}                          \affiliation{\Rochester}
\author{G.~Caceres}                       \affiliation{\CBPF}
\author{M.F.~Carneiro}                    \affiliation{\OregonState}
\author{D.~Coplowe}                       \affiliation{\oxford}
\author{H.~da~Motta}                      \affiliation{\CBPF}
\author{Zubair~Ahmad~Dar}                 \affiliation{\AMU}
\author{G.A.~D\'{i}az~}                   \affiliation{\Rochester}  \affiliation{\PUCP}
\author{J.~Felix}                         \affiliation{\Guanajuato}
\author{L.~Fields}                        \affiliation{\FNAL}  \affiliation{\Northwestern}
\author{A.~Filkins}                       \affiliation{\WM}
\author{R.~Fine}                          \affiliation{\Rochester}
\author{A.M.~Gago}                        \affiliation{\PUCP}
\author{H.~Gallagher}                     \affiliation{\Tufts}
\author{A.~Ghosh}                         \affiliation{\USM}  \affiliation{\CBPF}
\author{R.~Gran}                          \affiliation{\UMD}
\author{D.A.~Harris}                      \affiliation{\york}  \affiliation{\FNAL}
\author{S.~Henry}                         \affiliation{\Rochester}
\author{S.~Jena}                          \affiliation{\Mohali}
\author{D.Jena}                           \affiliation{\FNAL}
\author{J.~Kleykamp}                      \affiliation{\Rochester}
\author{M.~Kordosky}                      \affiliation{\WM}
\author{D.~Last}                          \affiliation{\upenn}
\author{T.~Le}                            \affiliation{\Tufts}  \affiliation{\Rutgers}
\author{A.~Lozano}                        \affiliation{\CBPF}
\author{E.~Maher}                         \affiliation{\MCLA}
\author{S.~Manly}                         \affiliation{\Rochester}
\author{W.A.~Mann}                        \affiliation{\Tufts}
\author{C.~Mauger}                        \affiliation{\upenn}
\author{K.S.~McFarland}                   \affiliation{\Rochester}
\author{B.~Messerly}                      \affiliation{\Pittsburgh}
\author{J.~Miller}                        \affiliation{\USM}
\author{J.G.~Morf\'{i}n}                  \affiliation{\FNAL}
\author{D.~Naples}                        \affiliation{\Pittsburgh}
\author{J.K.~Nelson}                      \affiliation{\WM}
\author{C.~Nguyen~}                       \affiliation{\Florida}
\author{A.~Norrick}                       \affiliation{\WM}
\author{Nuruzzaman}                       \affiliation{\Rutgers}  \affiliation{\USM}
\author{A.~Olivier}                       \affiliation{\Rochester}
\author{V.~Paolone}                       \affiliation{\Pittsburgh}
\author{G.N.~Perdue}                      \affiliation{\FNAL}  \affiliation{\Rochester}
\author{M.A.~Ram\'{i}rez}                 \affiliation{\Guanajuato}
\author{R.D.~Ransome}                     \affiliation{\Rutgers}
\author{D.~Ruterbories}                   \affiliation{\Rochester}
\author{H.~Schellman}                     \affiliation{\OregonState}  \affiliation{\Northwestern}
\author{C.J.~Solano~Salinas}              \affiliation{\UNI}
\author{H.~Su}                            \affiliation{\Pittsburgh}
\author{M.~Sultana}                       \affiliation{\Rochester}
\author{V.S.~Syrotenko}                   \affiliation{\Tufts}
\author{E.~Valencia}                      \affiliation{\WM}  \affiliation{\Guanajuato}
\author{D.~Wark}                          \affiliation{\oxford}
\author{A.~Weber}                         \affiliation{\oxford}
\author{M.Wospakrik}                      \affiliation{\Florida}
\author{B.~Yaeggy}                        \affiliation{\USM}
\author{L.~Zazueta}                       \affiliation{\WM}

\collaboration{The MINER$\nu$A Collaboration}\ \noaffiliation

\date{\today}

\begin{abstract}
We have measured new observables based on the final state kinematic imbalances in the mesonless production of $\nu_\mu+A\rightarrow\mu^-+p+X$ in the \minerva\ tracker. Components of the muon-proton momentum imbalances parallel ($\dpty$) and perpendicular($\dptx$) to the momentum transfer in the transverse plane are found to be sensitive to the nuclear effects such as Fermi motion, binding energy and non-QE contributions. The QE peak location in $\dpty$ is particularly sensitive to the binding energy. Differential cross sections are compared to predictions from different neutrino interaction models. The Fermi gas models presented in this study cannot simultaneously describe features such as QE peak location, width and the non-QE events contributing to the signal process. Correcting the \genie's binding energy implementation according to theory causes better agreement with data. Hints of proton left-right asymmetry are observed in $\dptx$. Better modeling of the binding energy can reduce bias in neutrino energy reconstruction and these observables can be applied in current and future experiments to better constrain nuclear effects.


\end{abstract}

\maketitle

\section{Introduction}

  Neutrino oscillation experiments measure the final state particles produced by neutrino-nucleus scattering processes. Models that accurately describe these interactions are crucial to reducing the uncertainties in the measurements of oscillation parameters. 
  
  Most neutrino-nucleus interactions are modeled through the impulse approximation (IA), where the probe sees the target nucleus as a collection of independent nucleons and the resulting particles then evolve independently. Important components of modeling in the IA picture include the initial state nucleon's energy-momentum distributions, the nuclear potentials, and the final state interactions (FSIs) that modify the kinematics of the final-state particles as they propagate through the nucleus. 
  
  The leptonic system provides energy to the hadronic side of the reaction to bring a bound nucleon on-shell and separate it from the remnant nucleus. Such energy is often loosely referred to as ``binding energy'', but Ref~\cite{Bodek:2018lmc} draws a distinction between the different energy parameters in neutrino models and how their effects depend on the implementation details. 
  
  In this paper we refer to the average energy transferred to the target nucleus to bring a bound nucleon inside the target onto the mass shell as the ``removal energy'', represented in this paper by $\epsilon^{N(P)}$ for the neutron (proton) initial state in neutrino (antineutrino) interactions. The energy associated with nuclear potentials is referred to as the nuclear potential energy. The combined effects of the ``\removalE'' and the nuclear potential energy is referred to as the ``\intE''. The \intE\ implementation in the IA picture is discussed in detail in Sec.~\ref{sec:IA}.

 For many neutrino experiments, particularly at low energies like \ttk, \microboone, and  the second oscillation maximum in \dune, incorrect treatment of the \intE\ may significantly bias the reconstructed neutrino energy and will alter the expected kinematics of final state nucleons.  Such effects are already a significant systematic in the measurement of $\delta m^2_{23}$ in the T2K experiment~\cite{Bodek:2018lmc,Abe:2018wpn}. 
  
 This paper will examine a set of new observable quantities that are sensitive to nuclear effects, and especially to the interaction energy implementation used in generators. The variables are extensions to the recent measurements of momentum imbalance in mesonless events with a muon and proton in the final state, here referred to as single transverse kinematic imbalance (\stki)~\cite{PhysRevC.94.015503} by the \minerva~\cite{Lu:2018stk} and \ttk~\cite{t2kstki} experiments. 
  
  The new observables are derived from the \stki\ observable $\dptv$. Specifically, we define $\dpty$ to be the projection of $\dptv$ along the transverse component of the leptonic momentum transfer, which is sensitive to the effects of the \intE.  We also report on the cross section in $\dptx$, the $\dptv$ projection normal to the neutrino-muon interaction plane. The \stki\ variables and their sensitivities to the \intE\ is discussed in greater details in Sec.~\ref{sec:stki} and Sec.~\ref{sec:sensitivities} respectively.
  
  These quantities provide neutrino oscillation experiments with a method to evaluate the validity of the \intE\ implemented in the interaction models. The \intE\ affects the reconstructed energy scale in the  simulated neutrino interaction. Evaluating the uncertainties in the neutrino energy scale based on an inaccurate implementation of the \intE\ will result in unnecessarily large systematics.
  
  We measure $\dptx$ and $\dpty$ using \minerva\ $\nu_{\mu}$-induced muon-proton mesonless interactions on hydrocarbon at $\langle E_\nu\rangle=3\gev$; this is the same data set as was studied previously in Ref~\cite{Lu:2018stk}.  The differential cross sections in these quantities are compared with \genie~\cite{GENIE}, \neut~5.40~\cite{NEUT}, \nuwro~\cite{NUWRO} and \gibuu~\cite{Buss:2011mx,Gallmeister:2016dnq} event generator predictions. The methodology is detailed in Sec.~\ref{sec:method} and the results are discussed in Sec.~\ref{sec:results}.
  
\section{Impulse Approximation\label{sec:IA}}
  
   \begin{figure}[tbp]
      \centering
      \includegraphics[width = .45\textwidth, trim={0cm 0cm 0cm 1cm},clip]{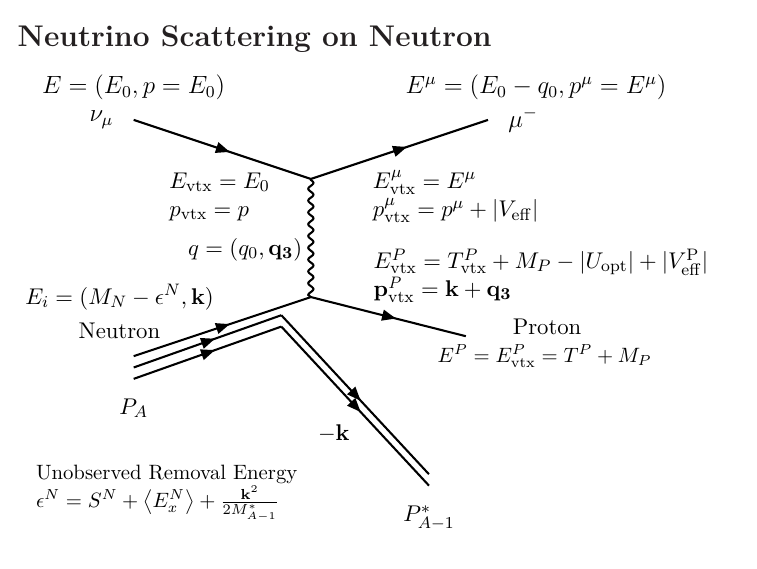}
      \caption{A neutrino interaction with a bound neutron in the impulse approximation. $\nu$, $\mu$, $N$ and $P$ are the neutrino, muon, neutron and protons respectively. The incoming neutrino with 4-momentum $E$ interacts with the bound neutron with 3-momentum $\vec{k}$ and \removalE\ $\epsilon^N$. The \removalE\ consists of the nucleon separation energy $S^N$, average excitation energy $\langle E_x^N\rangle$ and the kinetic energy of the remnant nucleus. $E^{P(\mu)}$ and $p^{P(\mu)}$ are the proton ($\mu$) total energy and momentum, $T^P$ is the proton kinetic energy,  $|\ufsi|$ and $|\veffp|$ ($|\veff|$) are the magnitudes of the optical potential and the Coulomb potential experienced by proton (muon). The quantities with the subscript ($\mathrm{vtx}$) are those immediately after 4-momentum transfer.   } 
  \label{fig:IA}
  \end{figure} 
 
  We illustrate the effects of \intE\ with the charged-current quasielastic (CCQE) interaction in the IA picture shown in Fig.~\ref{fig:IA}. In this picture, only a single nucleon is involved in the hard scattering, $\nu\neutron\to\mu^-\proton$.

  The neutrino with energy $E_0$ made 4-momentum transfer, $q=(\qzero,\vec{q_3})$ to a bound neutron of mass $M_N$ inside a target nucleus with $A$ nucleons, where $\qzero$ and $\vec{q_3}$ are the energy and momentum transfer respectively. The target nucleus was initially at rest with mass $\Ma$, the bound neutron has 4-momentum $E_i=(M_N-\epsilon^N, \vec{k})$ where $\vec{k}$ is neutron's Fermi momentum. The remnant nucleus with 4-momentum $P^*_{A-1}$ must have momentum $-\vec{k}$ for the target nucleus to be at rest. 
  The energy of the initial state neutron can be written as:
  \begin{equation}
      E_i = \Ma - \sqrt{\mas^2 + \vec{k}^2}, \label{eq:offshell}
  \end{equation}
  where $\mas$ is the mass of the excited spectator nucleus. For the nuclear targets typically used in neutrino experiments ($^{12}$C, $^{16}$O and $^{40}$Ar), we have $\mas^2\gg \vec{k}^2$. Then we can expand the initial state nucleon energy as:
  \begin{align}
      E_i &\approx \Ma - \mas-\frac{\vec{k}^2}{2\mas}\nonumber\\
          &= M_{N}-S^{N} - \ex^{N}-\frac{\vec{k}^2}{2\mas}.
          \label{eq:eiIA}
  \end{align} 
  
  The \removalE\ parameter 
  \begin{align}
      \epsilon^N = S^N+\ex^{N}+\left<T_{A-1}\right>,
      \label{eq:removalE}
  \end{align} 
  accounts for the \textit{neutron separation energy} from the target nucleus $S^N$,
  \begin{equation}
     S^N = M_{A-1} + M_N - M_A, \label{eq:S}\\
  \end{equation}
  \noindent and the excitation energy of the final state nucleus, $\ex^{N}$, when the initial state nucleon is a neutron,
  \begin{equation}
     \ex^{N} = M_{A-1}^{*} - M_{A-1}.\label{eq:Ex}
  \end{equation}
  The average kinetic energy $\left<T_{A-1}\right>=\vec{k}^2/(2\ma^*)$ of the excited remnant nucleus with A-1 nucleons affects the interaction only through its nuclear potentials. For neutrino QE interactions on $^{12}$C,  $S^N = 18.7\mev$, $\ex^{N}=10.1\mev$ and $\langle T_{A-1}\rangle = 1.4\mev$~\cite{Bodek:2018lmc}. The \removalE\ is the average energy needed to bring the neutron onto the mass-shell.
 
  There are additional effects associated with the nuclear potentials that should be accounted for. For example, the nuclear optical potential describes the nucleus as a medium with complex refractive index: the real part of the potential affects the allowed kinematics of the initial state lepton-nucleon system in the IA while the imaginary part is related to inelastic scattering as the outgoing nucleon is making an exit from the nucleus~\cite{Hodgson_1971}. Reference~\cite{Bodek:2018lmc}~ fits inclusive electron scattering data to determine the real part of the optical potential which depends on the 3-momentum of the outgoing nucleon at the interaction vertex.  This optical potential is denoted as $\ufsi[(\vec{k}+\vec{q}_3)^2]$ in this work. The effect of the optical potential is largest at lower momentum.  For carbon, the parameterization of Ref~\cite{Bodek:2018lmc} is 
  \begin{equation}
    \ufsi=\min \left[0,-29.1+(40.9/{\rm\textstyle GeV}^2)(\vec{k}+\vec{q}_3)^2)\right] \mev.
    \label{eq:ufsi}
  \end{equation}
  \noindent In this analysis we use this parameterization, and it is on average $2\mev$ for our selected events. 
  
  Another potential, the Coulomb potential $\veff$ of the positively charged remnant nucleus will modify the momenta of the outgoing charged particles as they propagate through the nucleus. In Fig.~\ref{fig:IA}, a distinction between the Coulomb potential experienced by muon ($|\veff|$) and proton ($|\veffp|$) is made; however,for neutrino interactions both particles experience the same Coulomb potential as the proton number in the nucleus remains unchanged after the interaction. 
  For carbon, $|\veff|$ is $3.1\mev$~\cite{Bodek:2018lmc}. \\
  
  Figure~\ref{fig:IA}  illustrates  energy and momentum conservation between the initial and final state. The total energy of the final proton and muon is equal to the total of the initial neutron and lepton, less energy required to create the final state excited nucleon in the reaction. The Coulomb potential affects any charged final state particles, but the optical potential affects the final state nucleon only. For example, the muon with total energy $E^\mu=E_\mathrm{vtx}^\mu$ begins inside the Coulomb potential, with kinetic energy $E^\mu+|\veff|$ and potential energy $-|\veff|$, and is decelerated during transport inside the nucleus medium so that its kinetic energy is $E^\mu$ outside the nucleus. The proton experiences both the Coulomb potential and the optical potential, which modify its kinetic energy and momentum but conserve the total energy.
  The full energy conservation equation on the hadronic side is as follows~\cite{Bodek:2018lmc}:
  \begin{multline}
      E_\mathrm{vtx}^P =\qzero + M_N - S^N - \exN - \frac{\langle \vec k^2\rangle}{2{M_{A-1}^*}^2} = \\
      \sqrt{\vec{(k+q_3)^2}+M_P^2} - |\ufsi[\vec{(k+q_3)}^2]| + |\veffp| \\
                   =E^{P}.
      \label{eq:1p1h}
  \end{multline}
  \noindent Here, the final state proton is assumed to be on-shell with energy $E_\mathrm{vtx}^P=E_f^{P}$,before and after exiting the nucleus. Its kinetic energy immediately after the 4-momentum transfer is
  \begin{equation}
      T_\mathrm{vtx}^P = \qzero + M_N - M_P -\epsilon^N,
  \end{equation}
  and is modified by the nuclear potentials so that outside the nucleus the kinetic energy becomes
  \begin{equation}
      T^{P}=T_\mathrm{vtx}^P-|\ufsi|+|\veffp|. 
  \end{equation}
  
  The \removalE\ used by neutrino Monte Carlo (MC) generators, such as \genie~\cite{GENIE}, \neut~\cite{NEUT}, and \nuwro~\cite{NUWRO}, are discussed in detail in Ref~\cite{Bodek:2018lmc}. These generators use variants of spectral functions, mostly the Fermi gas model in the IA picture with \removalE\ constrained by inclusive electron scattering data~\cite{Smith:1972xh}. However, they have distinct implementations of the IA model which affects the energy terms going into the \removalE\ parameter. For example, in \genie's IA implementation, the off-shell bound initial nucleon is generated with Eq.(\ref{eq:1p1h}), but with $E_x^N$, $\ufsi$, and $\veffp$ set to 0. 
  \genie\ subtracts an additional ``binding energy'' parameter  $\dN{nucleus}$ from the final state protons in QE processes to account for the \removalE. The implementation of this term is independent of the kinematics at the interaction vertex, which causes the energy of the final state nucleons to be biased.
  The values of $\dN{nucleus}$ were measured by Ref~\cite{Smith:1972xh} and referred to as the ``Moniz interaction energy'' in Ref~\cite{Bodek:2018lmc}. The Moniz interaction energy is an empirical fit to the sum of the \removalE\ and the nuclear potentials, but for a non-relativistic on-shell formalism. For $\nu+^{12}$C scattering, $\dN{C} = 25~\mev$~\cite{GENIE}. 
  
  \begin{table*}[!tbh]
    \caption{Calculated energy corrections to the final state leptons and hadrons from the \genie\ generator for QE neutrino scattering on $^{12}C$, $\dN{C}=25\mev$, $E_x=10.1\mev$. Other interaction channels are not altered. }
    \label{tab:GenieCorrectionModes}
     \begin{tabular}{|l|c|c|c|c|}
      \hline
      Correction    & $E^P = E^P_\genie+\delta^P$ & $E^\mu = E^\mu_\genie+\delta^\mu$  & \genie~Baseline Shift & QE Baseline Shift\\
        &   $\delta^P~(\mev)$  & $\delta^\mu~(\mev)$& $\left<\delta^P\right>$,$\left<\delta^\mu\right>$ $(\mev)$ &  $\langle\dpty\rangle (\mevc)$    \\ \hline
      0: Default               &  0                    & 0         & 0,0    & 0            \\ 
         (no corrections)                       &                       &           &       &               \\ \hhline{|=|=|=|=|=|}
      
      1: $\ufsi$ only  & $\dN{C} - |\ufsi|$    & $|\ufsi| - E_x$ & $22.7,-7.8$ & $29.4$    \\
      (w/ $E_x\&\dN{C}$) &    &  &  &      \\ \hline
      2: $\ufsi$ and $\veff$ & $\dN{C} - |\ufsi|$    & $|\ufsi| - E_x$ & $25.8,-10.9$ & $33.9$\\
       (w/ $E_x\&\dN{C}$)& $+|\veffp|$    & $-|\veff|$ & &\\ \hline
    \end{tabular}
\end{table*}
  
  In this paper, we refer to the collective energy shifts due to \removalE\ and the nuclear potentials as the ``\intE'', in the spirit of the Moniz interaction energy of Ref~\cite{Bodek:2018lmc}. This \intE\ is specific to the off-shell formalism described in Eq.~\ref{eq:1p1h}.
  
  We simulate the effects of the \intE\ implementations in \genie\ by modifying the final state muon and proton energies after a sample is generated according to Table~\ref{tab:GenieCorrectionModes}.  The corrections outlined are motivated by the study in Ref~\cite{Bodek:2018lmc}. 
  Comparisons between the default \genie\ implementation ( 0 in Table~\ref{tab:GenieCorrectionModes}) and two different corrections (1 and 2) are made. For both sets of corrections, which are applied to QE events, we add $\dN{C}$ back to the exiting proton to undo the bias, we then subtract $E_x$ from the muon to account for the shift in momentum transfer in the leptons (derived in Appendix~\ref{app:genieCorr}). In addition, the correction 1 applies an optical potential correction to both the muon and proton, while the correction 2 applies the Coulomb correction on top of correction 1.  The average $|\ufsi|$ is $\approx 2\mev$ for the proton and muon kinematics chosen.  
  
  The corrections are approximations which implement the leading effect of the nuclear potential.  That potential will also cause changes, small for our events that have an energetic final state proton, in the four momentum transferred to the off-shell target nucleon.  Appendix~\ref{app:genieCorr} provides the derivation of our corrections.

\section{The \stki\ variables\label{sec:stki}}
 
  \begin{figure}[tbp]
      \centering
      \includegraphics[width=0.5\textwidth]{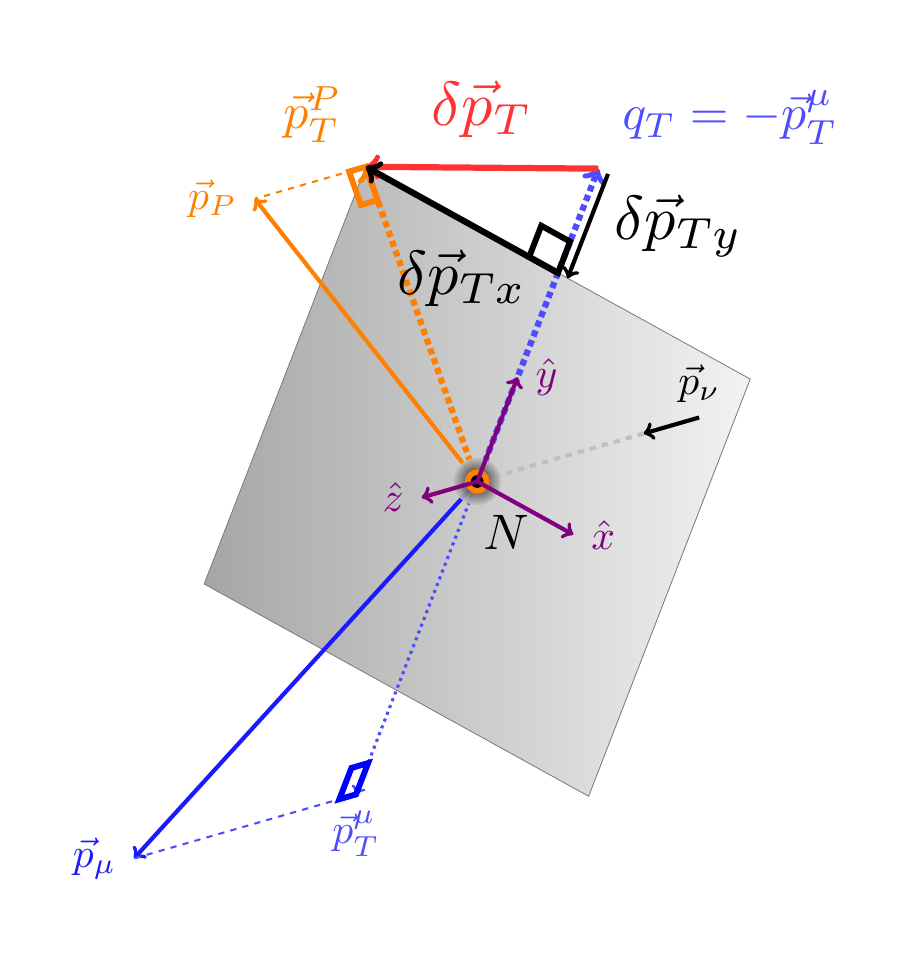}
      \caption{Schematics of the Single Transverse Kinematic Imbalances and their projections.The incoming neutrino interacts on the neutron ($N$) in the nucleus. The neutrino direction $\pnuh$ forms the $z$ axis. A final state muon with $\vec{p}_\mu$ and a proton with $\vec{p}_p$ are produced. The muon transverse momentum is $\ptl$, and $-\ptlh$ defines the $y$ axis. The proton transverse momentum is $\ptni$ and decomposed along $x$ and $y$ axis respectively. In this example, both $\dptx$ and $\dpty$ are negative, and only the distribution of $\dptx$ for QE events is expected to be symmetric around zero.}
      \label{fig:kinematic_imbalance}
  \end{figure}
    \begin{figure}[tbp]
    \centering
    \includegraphics[width=.45\textwidth]{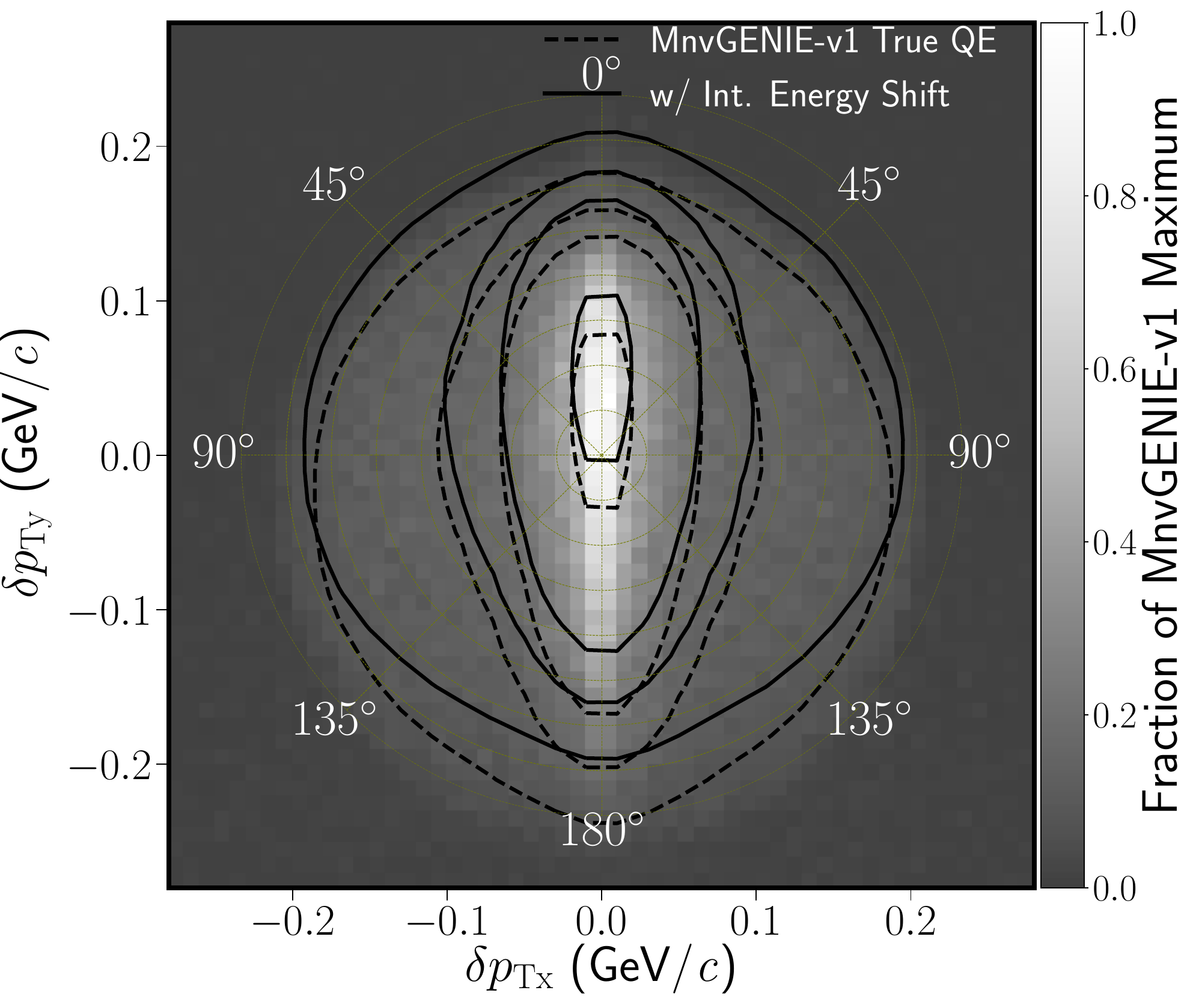}
    \caption{ $\dpty$ vs.\ $\dptx$ for CCQE events in \genie. The contours, from outside towards the center, represent 0.1, 0.2, 0.3, and 0.7 of maximum.  The angles correspond to $\dalphat$ values. The neutrino direction points out of the page.  The dashed contour describes the default \genie~distribution with a \mnvgenie{1}\ tune, described in Sec.~\ref{sec:method}}. The solid contour shows the shift in the distribution after correction is made to the \intE\ with correction 01 of Table~\ref{tab:GenieCorrectionModes}. There is negligible deformation in the $\dptx$ direction compared to the shift in $\dpty$. 
    
    \label{fig:contour}
  \end{figure}

  The \stki~measurements for CCQE-like events, which include a lepton, at least one proton and no mesons in the final state, are defined in Ref~\cite{PhysRevC.94.015503}:
  \begin{align}
      &\dptv \equiv \ptni+\ptl,\\
      &\dalphat\equiv\arccos\left(-\ptlh\cdot\dptvh\right),
  \end{align}
\noindent
  where $\ptni$ and $\ptl$ are the components of proton and muon momenta in the plane perpendicular to the neutrino direction. The \stki~variable $\dptv$ and its decompositions along the Cartesian coordinate system defined with respect to the neutrino and muon kinematics are illustrated in Fig.~\ref{fig:kinematic_imbalance} and mathematically defined as
  \begin{align}
      &\dptx = \left(\pnuh\times\ptlh\right)\cdot \dptv,\nonumber\\
      &\dpty = -\ptlh\cdot\dptv.
      \label{eq:dptxy}
  \end{align}
  Here, $\pnuh$ is the neutrino direction, $\dpty$ is anti-parallel to the muon transverse direction $\ptlh$ while $\dptx$ is perpendicular to $\dpty$ along the normal of neutrino-muon plane. The coordinate system describing $\dptx$ and $\dpty$ is relative to the neutrino and muon kinematics. Specifically $\vec{\hat{y}}$ is along the transverse component of 3-momentum transfer, $\vec{\hat{z}}$ is along the neutrino direction and there is no 3-momentum transfer in the $\vec{\hat{x}}$ direction. Both $\dptx$ and $\dpty$ can be measured from the final state particles. Any \intE\ effect will mostly affect the 4-momentum transfer and $\dpty$. For CCQE events, $\dptx$ is expected to symmetrically distribute on both sides of the neutrino-muon interaction plane.

  $(\dptx,\dpty)$ can be defined in terms of $(\dpt,\dalphat)$ as:
  \begin{align}
      |\dptx|&=\dpt\sin\dalphat,\nonumber\\
      \dpty&=\dpt\cos\dalphat.
      \label{eq:relation}
  \end{align} 
  Here $\dpty$ is positive if the proton has gained momentum along $-\ptl$. 
  Figure~\ref{fig:contour} illustrates the relationship between $(\dptx, \dpty)$ and $(\dpt, \dalphat)$ as the different projections of $\dptv$ in the Cartesian and the polar coordinate systems respectively. The resulting distribution in the $\dalphat$ and $\dpt$ residuals provides insights into other nuclear effects affecting the cross section, such as FSIs, the Fermi motion and two-particle-two-hole (2p2h) processes\cite{Lu:2018stk}.

\section{Sensitivities to \intE\ implementation \label{sec:sensitivities}}
  The shapes of $\dptx$ and $\dpty$ are affected by nuclear effects. The non-zero width of $\dptx$ for the QE portion of the signal is largely due to the Fermi motion of the target nucleus.  The average Fermi momentum in Carbon is approximately $221\mev$.  In the absence of FSI effects, this is the only momentum available in the $x$ direction. FSIs could alter the outgoing protons' directions, but in Carbon, an outgoing nucleon typically exits without interacting with the nucleus, or interacts with the nucleus elastically producing a small change in direction.

  The momentum transferred to the hadronic system is confined in the $yz$ plane.  On an event by event basis, the nuclear potential may alter this momentum as well, and therefore the direction of the final state nucleon, but this effect averages to zero because the initial state nuclear momentum is on average zero.  Therefore, changes to the \intE\ at the event vertex, on average, only alter $\dpty$.  Mathematically, the effect of the \intE\ is as follows:
  
  For an outgoing nucleon with energy $E_f^\prime$ before it has left the region of nuclear potentials, its momentum $\vec{p}_f$ as a function of an energy shift $\tau$ due to the \intE\ is:
  \begin{align}
   |\vec{p}_f(\tau)|= \sqrt{ (E_f' - \tau)^2 - M_P^2 },
   \end{align}
   where $E_f' =\sqrt{M_P^2+\vec{p}_0^2}$ and $\vec{p}_0 = \vec{p}_f(0)=\vec{k}+\vec{q}$. In the limit 
   \begin{align}
       \frac{\tau E_f'}{p_0^2}\ll 1,
   \end{align}
   we can approximate $\vec{p}_f(\tau)$ by
   \begin{align}
   \vec{p}_f(\tau) \approx \left(1-\frac{E_f'}{p_0^2}\tau\right)\vec{p}_0.
  \end{align}
  Defining $\alpha=\tau E_f'/p_0^2$, we can write the 4-momentum conservation equation without FSI as:
    \begin{align}
        \begin{pmatrix}
        \qzero \\ 0 \\ q_T \\ q_L
        \end{pmatrix} 
        +
        \begin{pmatrix}
        E_i \\ k_x \\ k_y \\ k_z
        \end{pmatrix} 
        \approx    
        \begin{pmatrix}
        E_f' \\ p_{0x} \\p_{0y} \\ p_{0z}
        \end{pmatrix}-
        \begin{pmatrix}
        \tau \\ \alpha p_{0x} \\\alpha p_{0y} \\ \alpha p_{0z}
        \end{pmatrix},
        \label{eq:4mom}
    \end{align}
    where $(0,q_T,q_L)$ are components of the 3-momentum transfer $\vec{q}$, $(k_x,k_y,k_z)$ are components of Fermi motion $\vec{k}$. In this picture, $q_T$ is directly measurable as the transverse component of muon momentum, with magnitude $\ptlm$, but $q_L$ cannot be directly measured and estimates depend on the model used to calculate neutrino energy.
   
    The transverse components of the 3-momentum imbalance are
    \begin{align}
     \dptx&= (1-\alpha)p_{0x} \nonumber\\
     &\approx k_x-\alpha p_{0x}=k_x- \tau\frac{E_f}{p_0^2}p_{0x},\\
     \dpty &= (1-\alpha)p_{0y}+\vec{p_{\mu}}\cdot \hat{y}\nonumber\\
     &=p_{0y}-p_T^{\mu}-\alpha p_{0y}=p_{0y}-q_T-\alpha p_{0y}\nonumber\\
     &\approx k_y-\alpha p_{0y}=k_y-\tau\frac{E_f}{p_0^2}p_{0y},
      \label{eq:p0xy}
    \end{align}
    where we have assumed $k_x\approx p_{0x}$ and $q_T+k_y\approx p_{0y}$ because the Fermi momentum is large compared to the \intE-induced change in momentum.
    In the limit $\tau, \alpha \to 0$, $\dptx$ and $\dpty$ are the transverse components of the Fermi momentum, $(k_x,k_y)$.  The effect of energy shift, $\tau$, in each component of $(\dptx,\dpty)$ is then proportional to that component of $p_{0}$.  When $p_T^{\mu}\gg k_y$, the shift in $\dpty$ will be larger than the shift in $\dptx$.  In both components, the \intE\ effects acting on the Fermi momentum will average to zero, whereas the effects on $\dpty$ from $q_T$ will yield a net average shift in $\dpty$. For events in \genie\gvernew, there is approximately $+15\mevc$ offset in $\dpty$.
    The last column in Table~\ref{tab:GenieCorrectionModes} shows how applying energy corrections to the final state proton affects the average QE peak positions in $\dpty$. 
    \begin{figure*}[tbp]
      \centering
      \includegraphics[width=.45\textwidth,trim={0 .2cm 0 1.0cm},clip]{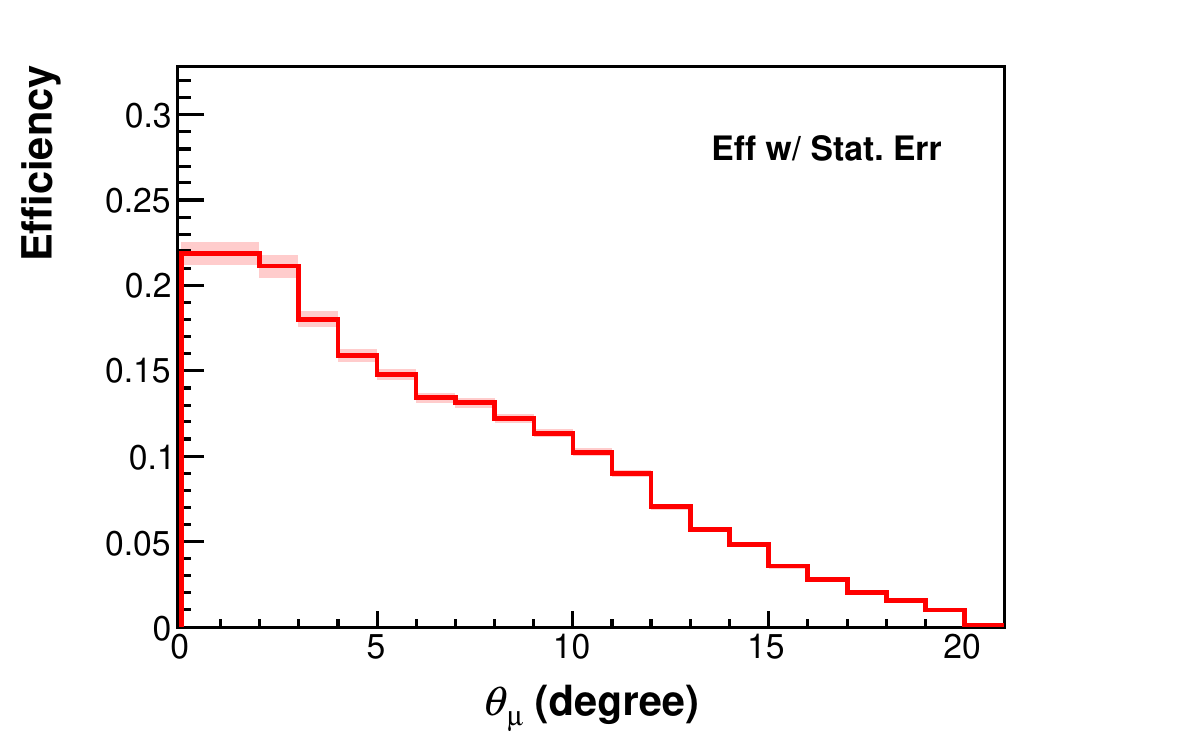}
      \includegraphics[width=.45\textwidth,trim={0 .2cm 0 1.0cm},clip]{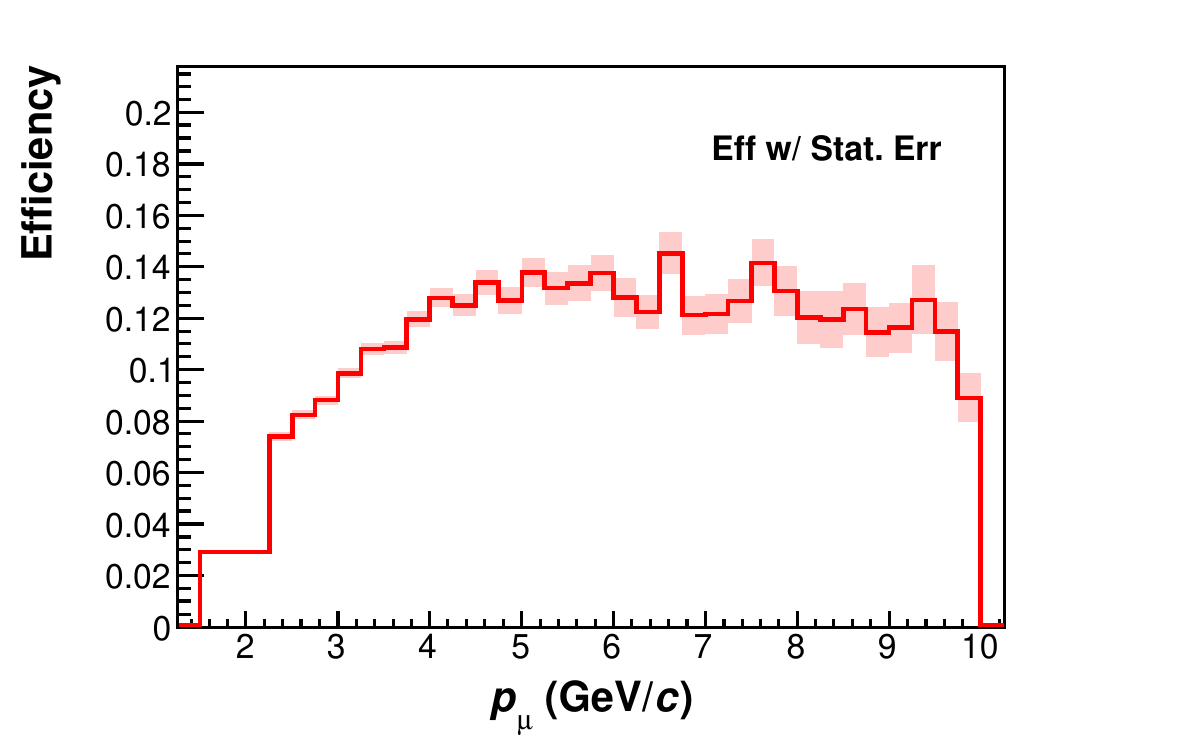}\\
      \includegraphics[width=.45\textwidth,trim={0 .2cm 0 1.0cm},clip]{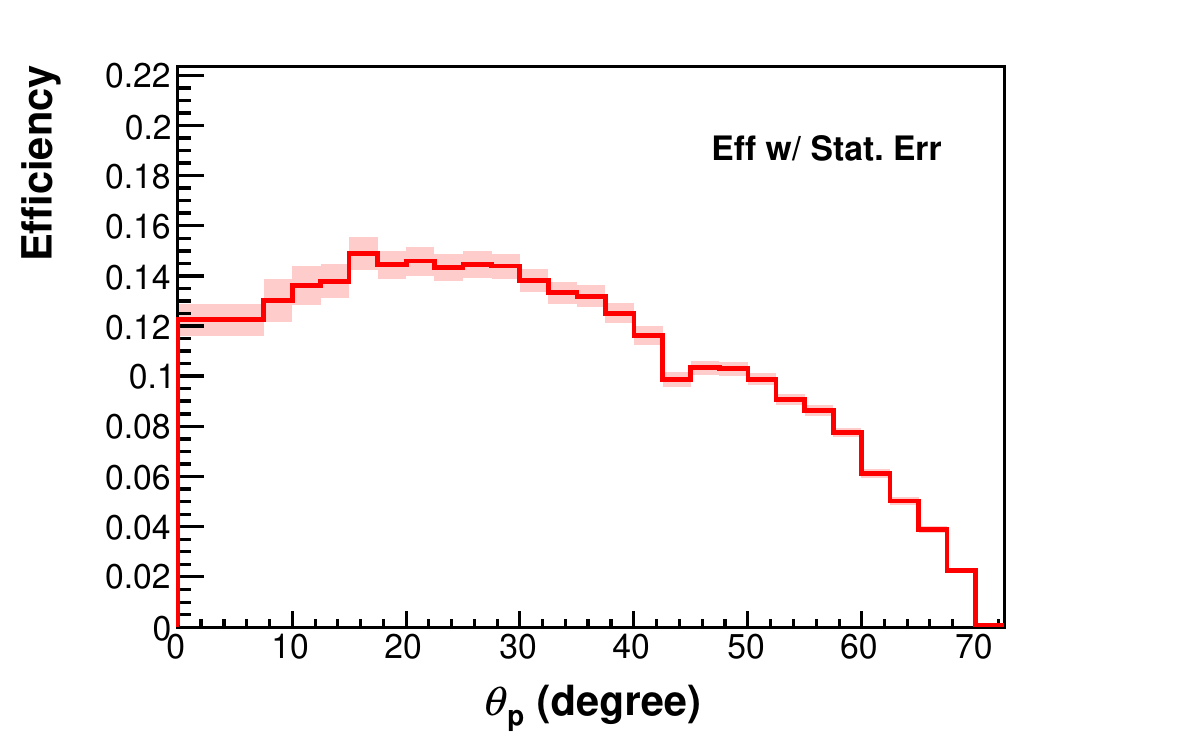}
      \includegraphics[width=.45\textwidth,trim={0 .2cm 0 1.0cm},clip]{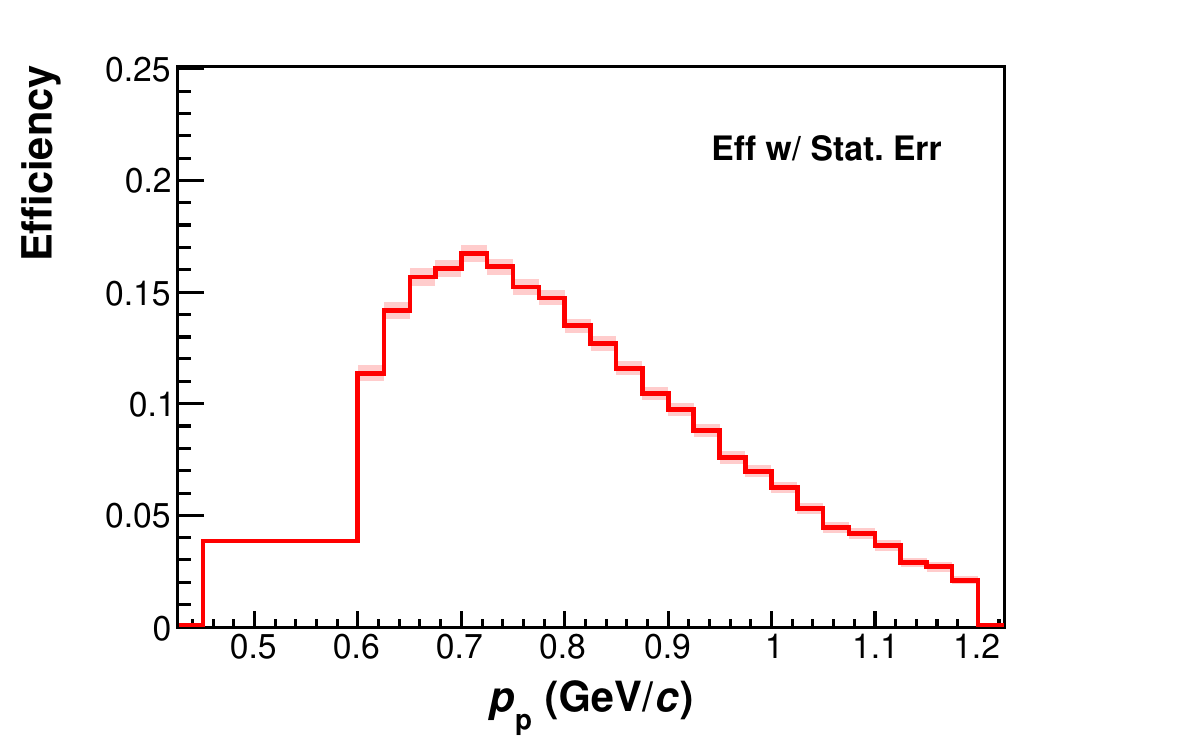}\\
      \caption{Muon (top) and proton (bottom) efficiencies as functions of opening angle $\theta$ (left) and momentum (right).}
      \label{fig:fsefficiencies}
  \end{figure*}
  \begin{figure*}[tp]
  \includegraphics[width=.45\textwidth,trim={0 .3cm 0 1.2cm},clip]{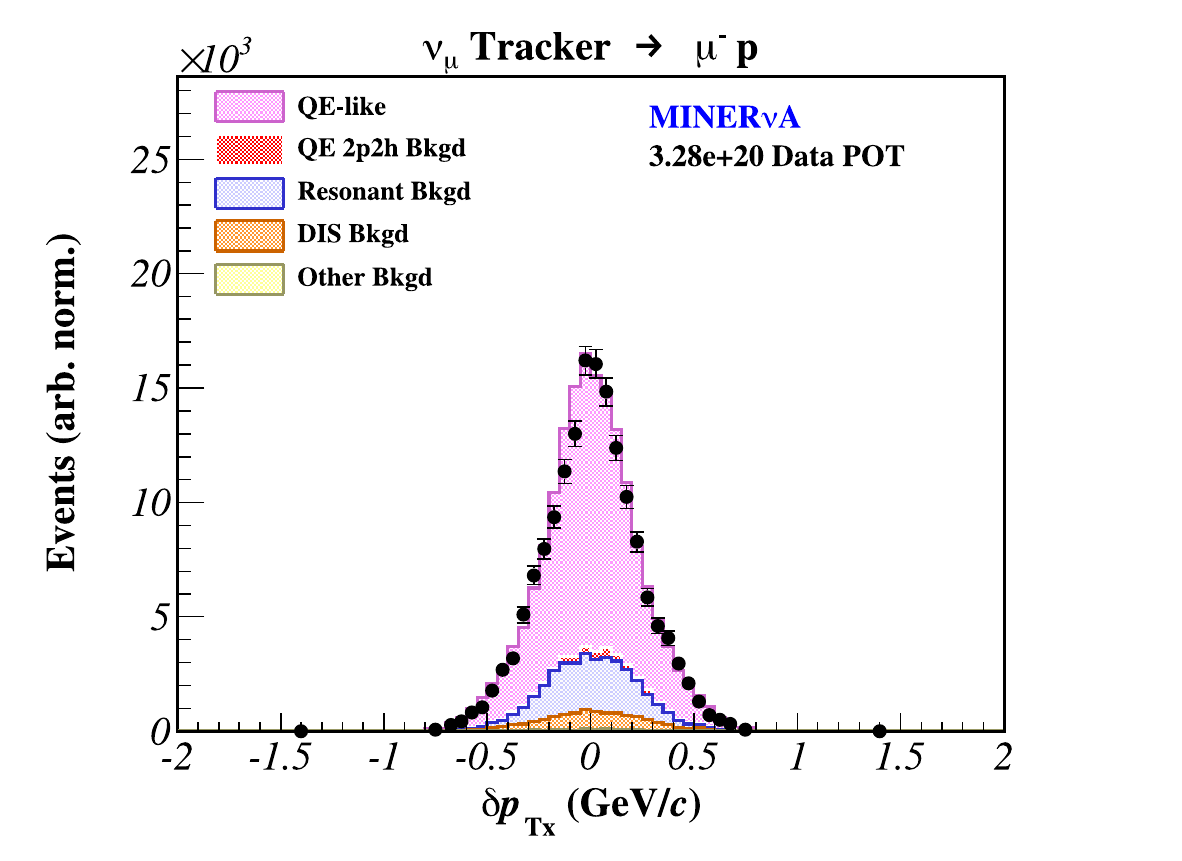}
  \includegraphics[width=.45\textwidth,trim={0 .3cm 0 1.2cm},clip]{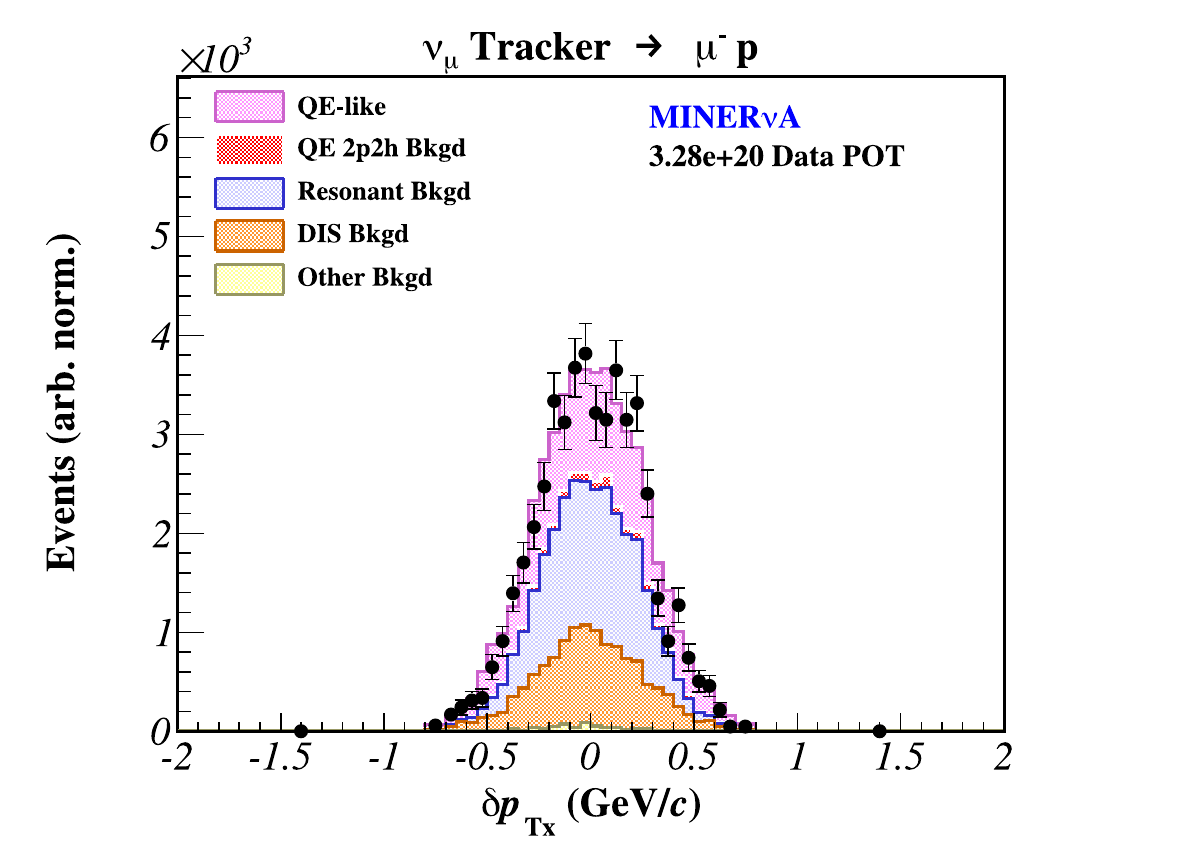}\\
  \includegraphics[width=.45\textwidth,trim={0 .3cm 0 1.2cm},clip]{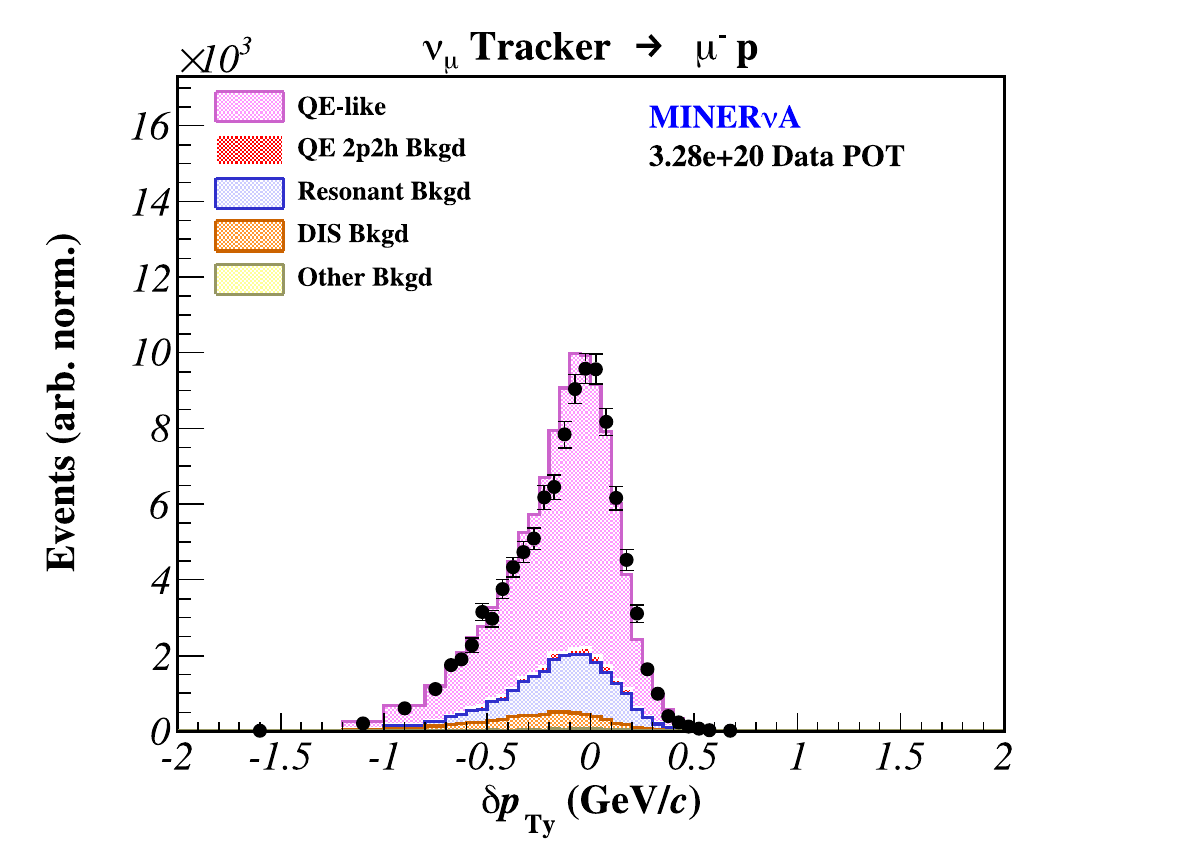}
  \includegraphics[width=.45\textwidth,trim={0 .3cm 0 1.2cm},clip]{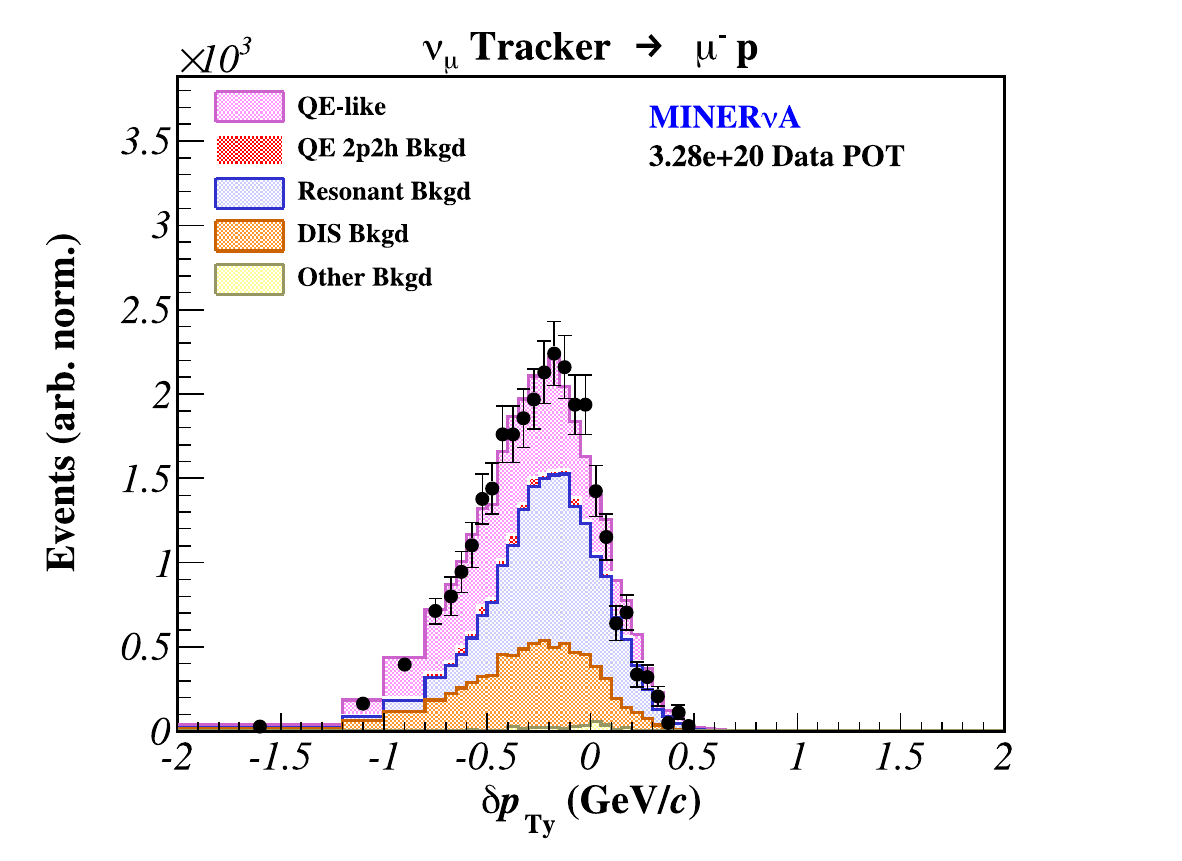}\\
  \caption{Reconstructed event rate in the $\dptx$ signal (top left) and a representative background sideband (top right); the $\dpty$ signal (bottom left) and a background sideband (bottom right). The background fraction in the signal have been fitted with a data-driven constraint using the sidebands. }
  \label{fig:evtrate_dptxy}
  \end{figure*}
\section{Apparatus and Methodology\label{sec:method}}

  The measurements of differential cross sections in $\dptx$ and $\dpty$ with the \minerva\ detector~\cite{Aliaga:2013uqz} use the same sample and methodology of the measurements described in Ref~\cite{Lu:2018stk}.  
  The signal requires no pions, one muon, and at least one proton in the final state, satisfying
    \begin{empheq}[left=\empheqlbrace]{align}
    1.5~\gevc<&\pmu<10~\gevc,~\thetamu<20^\circ,\label{eq:pmuon}\\
    0.45~\gevc<&\pproton<1.2~\gevc,~\thetaproton<70^\circ,\label{eq:thetaproton}
    \end{empheq}
  where $\pmu$ and $\thetamu$ ($\pproton$ and $\thetaproton$) are the final-state muon (proton) momentum and opening angle with respect to the neutrino direction, respectively. The data set  corresponds to 3.28$\times10^{20}$ protons on target (POT) delivered between 2010 and 2012 by the NuMI beam line~\cite{Adamson:2015dkw} at Fermilab. For this beam, the integrated $\nu_\mu$ flux is predicted to be $2.88\times10^{-8}/\textrm{cm}^2/\textrm{POT}$\cite{Aliaga:2016oaz}.

  Neutrino interactions are simulated with \genie~\gver~\cite{GENIE} in both a nominal form, and also with a \minerva ``tune'' (\mnvgenie{1.0.1}). The nominal \genie~ generates initial states with a modified Fermi Gas model containing contributions from the Bodek-Ritchie tail~\cite{PhysRevD.23.1070}. The CCQE cross-section is produced by the Llewellyn Smith formalism~\cite{LlewellynSmith:1971uhs}, with a dipole axial form factor with axial mass $M_A^{QE} = 0.99\gevcc$. Resonant pion production is modeled by the Rein-Sehgal~\cite{REIN198179} model.  Deep inelastic scattering is simulated with a Quark-Parton Model parameterized with the Bodek-Yang structure functions~\cite{Bodek:2002vp}. FSI is simulated with the \genie~hA model. 
  
  The tuning is based on \mnvgenie{1}, which has been applied in previous publications \cite{Lu:2018stk,Ruterbories:2018gub,Patrick:2018gvi}. \mnvgenie{1}~includes the Valencia two-particle-two-holes (2p2h) model~\cite{Nieves:2013rja,PhysRevD.88.113007, Schwehr:2016pvn} for two-body current simulation. Furthermore, the interaction strength of this 2p2h model has been tuned to \minerva~inclusive scattering data~\cite{PhysRevLett.116.071802}, resulting in a significant enhancement relative to the Valencia model in a restricted region of energy-momentum transfer. \mnvgenie{1}~also includes a non-resonant pion reduction to 43\% of the nominal as constrained by comparisons with bubble chamber deuterium data~\cite{Wilkinson:2014yfa,Rodrigues:2016xjj}. There is also a modification to the collective excitations of the nucleus for the CCQE channel, approximated as a superposition of 1p1h excitations and calculated with the Random Phase Approximation (RPA) in Ref~\cite{Nieves:2004wx} and uncertainties in Ref~\cite{Gran:2017psn}. The effects of non-resonant pion production and RPA in this analysis are negligible.
  
  On top of the \mnvgenie{1} tuning, \mnvgenie{1.0.1} removes QE events with elastic nucleon-nucleus FSI, replacing them with events where there is no FSI, to remove the effect of a mistake in \genie\'s implementation of the elastic nucleon-nucleus FSI.  The primary effect in the final state is in the angular distribution of outgoing protons.  A detailed discussion of this mistake can be found in Appendix~\ref{app:elasticFSI}.
  
  Reconstructed events with one  muon and at least one proton in the \minerva\ tracker satisfying Eq.(\ref{eq:pmuon})-(\ref{eq:thetaproton}) are selected. 
  Figure~\ref{fig:fsefficiencies} shows the reconstruction efficiencies of the muons and the protons due to event selection and detector acceptance.

  Only the muons which exit from the back of the \minerva\ detector and end up in the \minos\ detector can be fully reconstructed. The muon momentum lost inside \minerva\ is measured by energy deposits. The momentum in \minos\ is estimated by range or curvature, which depends on whether the muon is contained in the \minos\ spectrometer. 
  
  Proton identification is done with a track-based $\dEdx$ algorithm which could reconstruct the proton energy (including rescattered protons) to $5\%$ energy resolution\cite{Walton:2013es}. 
  An additional $\dEdx$ selection is applied on these protons to favor ones that interact elastically and contained (ESC) within the CH tracker, which improves the energy resolution to $3\%$~\cite{Lu:2018stk,Lu:2016mjf}. The ESC requirement impacts the selection efficiencies for protons with higher momentum, which tend to rescatter inelastically more often.
  
  The reconstructed proton energy and angular resolutions are $3\%$ and $2^\circ$, while the reconstructed muon energy and angular resolutions are $\sim 8\%$ and $0.6^\circ$. 
  The resolutions of the composite variables $\dptx$ and $\dpty$ have been evaluated to be $0.05\gevc$ and $0.06\gevc$ respectively.
  
  After the event selection, background contributions are estimated using predictions from \genie~\gver.
  The predicted background consists of events with pions in the final states, which mostly comes from RES and DIS interaction channels. The background is then constrained with a data-driven method with sidebands described in Ref~\cite{Walton:2014dka}. The event rate in the signal region and in a representative sideband for $\dptx$ and $\dpty$ are shown in Fig.~\ref{fig:evtrate_dptxy}. 
  In this figure, the sideband sample shown contains events with off-track visible recoil energy between $0.06$ and $0.385\gev$. Four sidebands with different visible recoil energy are used to constrain the backgrounds in bins of proton $Q^2_\mathrm{QE}$ from $0.15$ to $2.0\gev^2$. Separate weights are used for inelastic events with a baryon resonance events and for other, higher $W$, inelastic backgrounds.

  After subtracting the fitted background, the signal fraction is treated with an iterative unfolding procedure~\cite{DAgostini:1994fjx} to account for the detector resolution~\cite{Lu:2018stk}. Four iterations are chosen~\cite{Lu:2018stk} to balance model bias and statistical uncertainties in the unfolded distribution. The stability of the unfolding with four iterations is studied by unfolding different pseudodata sets with model variations different from our assumed cross section model.   As an extreme test, one of the variations we study for each of $\dptx$ and $\dpty$ puts in a large, non-physical, asymmetry in the relevant distribution.  For each of these pseudodata studies, we compare the consistency of the unfolded pseudodata with the input model assumption as a function of number of iterations.  For each pseudodata set, statistical uncertainties are added about the mean data prediction from the mode variation.  One thousand pseudodata sets are created for each study. We find that four iterations of unfolding are sufficient to achieve good agreement, where the metric for agreement is the mean $\chi^2$ from the comparisons of unfolded pseudodata to its true distribution.  We also verify that the mean $\chi^2$ fails to decrease significantly with additional iterations.
  
  The unfolded data is corrected for the predicted efficiency calculated as ratio between the predicted and selected number of simulated events in each bin. 
  The efficiencies for $\dptx$ and $\dpty$ in the QE region $\range{-0.5}{0.5}\gevc$ are constant at $0.1$ with $10\%$ relative variations and slowly falls by a factor of 2 over the regions $\pm\range{0.5}{0.1}\gevc$. The flux-averaged differential cross sections are then obtained by normalizing the efficiency-corrected distribution with the number of target nucleons ($3.11\times10^{30}$) and the predicted $\nu_\mu$ flux.
  
\begin{figure}[bp]
      \includegraphics[width=.5\textwidth,trim={0 0 0 0.21cm},clip]{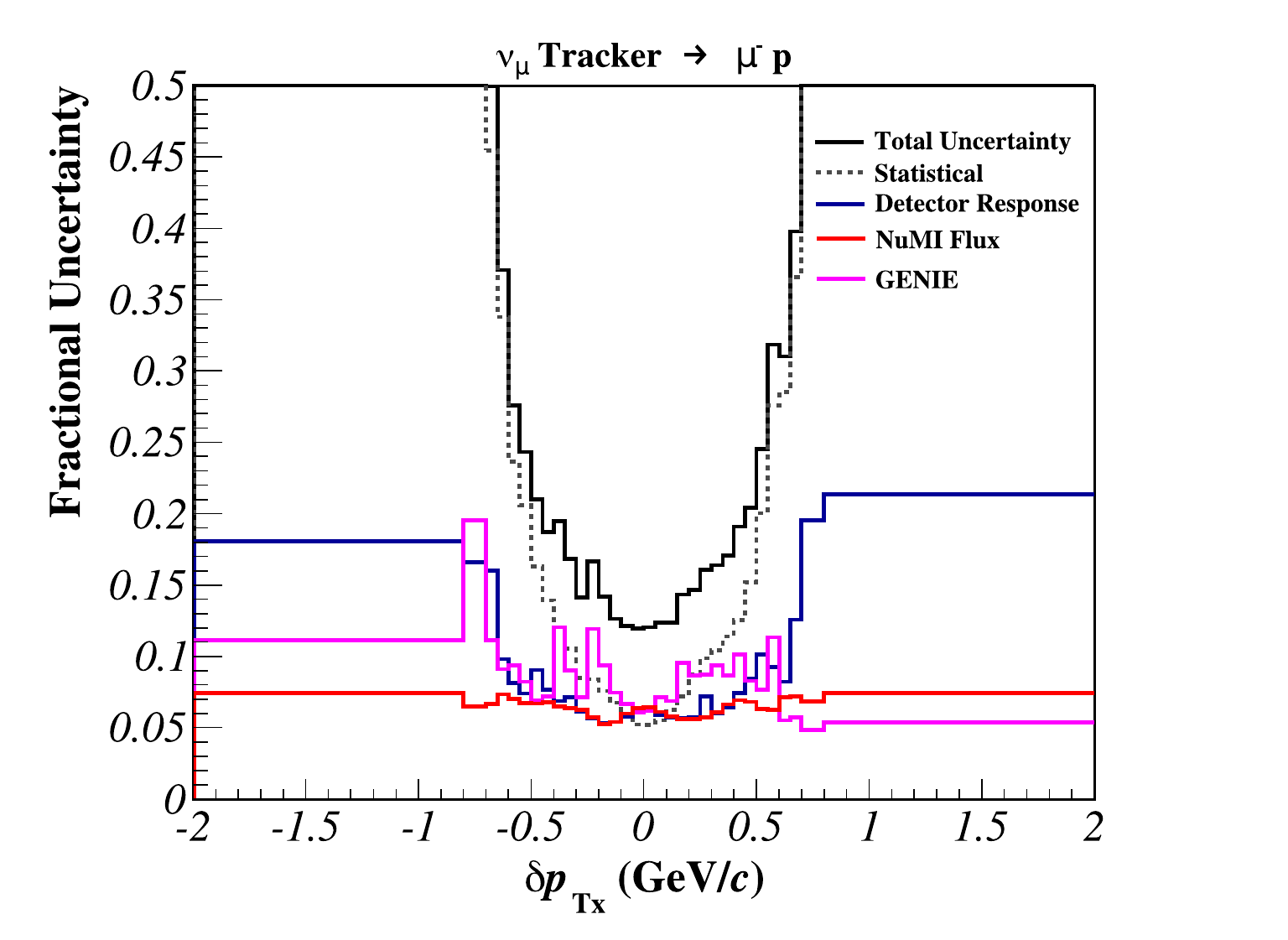}
      \includegraphics[width=.5\textwidth,trim={0 0 0 0.21cm},clip]{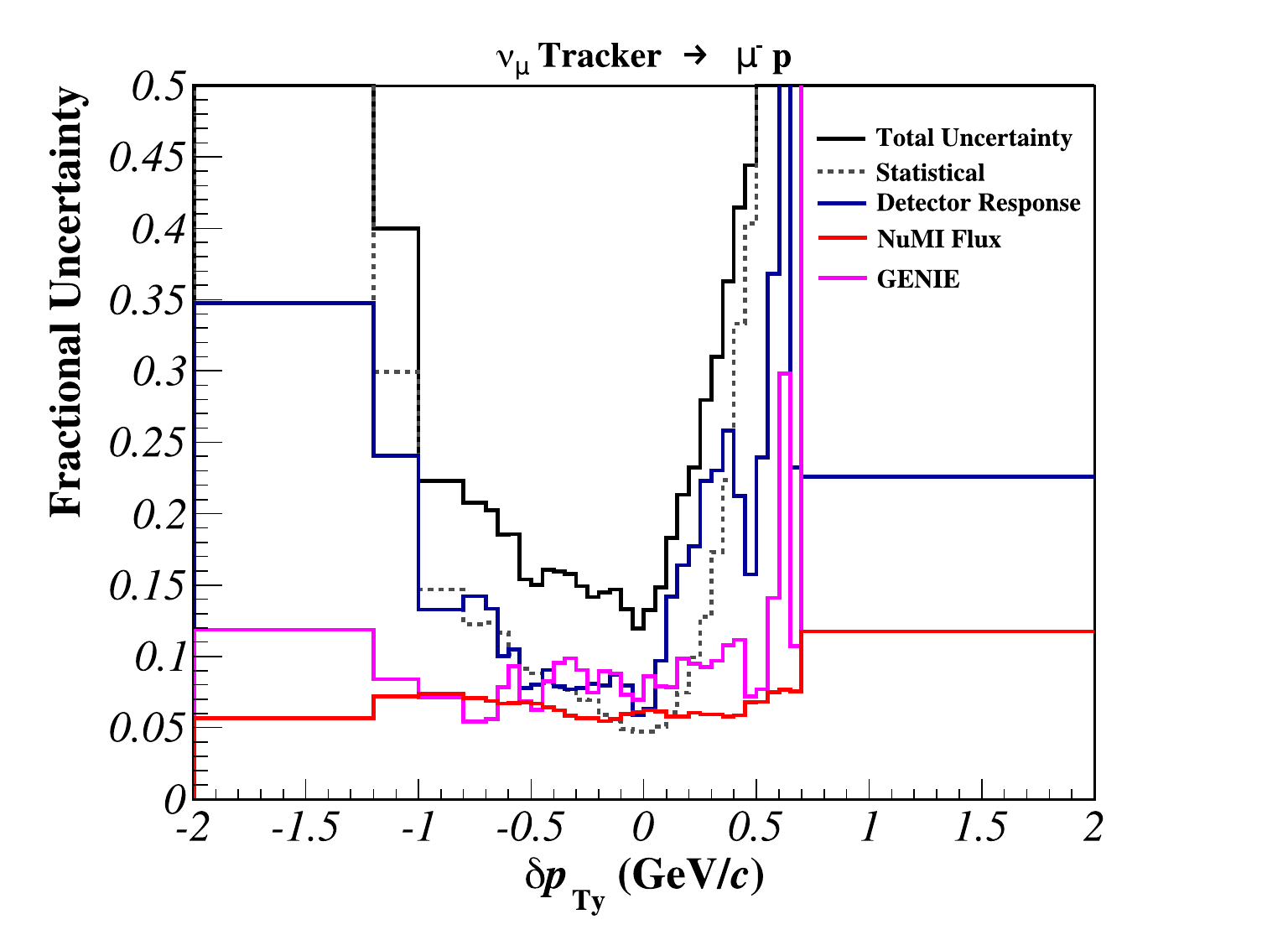}
      \caption{ Uncertainties on the extracted $\dptx$(top) and $\dpty$(bottom) cross sections.}
      \label{fig:uncertainties}
  \end{figure}

  Uncertainties on $\dptx$($\dpty$) result from statistical fluctuations and uncertainties in the NuMI flux prediction, the \genie\ modelling, and the detector response. The uncertainties are propagated throughout the cross section extraction procedure, and the results are summarized in Fig.~\ref{fig:uncertainties}. 

  The final differential cross sections in $\dptx$ and $\dpty$ are reported over $-0.7\gevc<\dptx,\dpty<0.7\gevc$.  Each broad category of systematic uncertainties, neutrino flux, detector response, and assumed interaction model (``\genie'') ranges between $5\%$ to $10\%$ within this region.  The largest contributing factor to uncertainty in the detector response is the tracking efficiency; the largest uncertainty in the neutrino interaction model is \genie's model of pion absorption in final state interactions.

\section{Results and Discussions\label{sec:results}}
 
  Model comparison is facilitated with the \nuisance~\cite{Stowell:2016jfr} neutrino interaction cross section comparison package. For the primary comparison with data, we use \genie~\gvernew~ with the Valencia 2p2h model replacing the default empirical 2p2h model. \nuisance\ is used to apply the \mnvgenie{1.0.1}\ tune that is described above. \genie~\gvernew\ and \genie~\gver\ have consistent model implementations. A careful internal \minerva\ study indicates the main difference for this analysis is an increase of $S^N$ by $14.8\mev$ from changes to the nuclear masses in \genie.
  
   \begin{figure}[tbp]
    \includegraphics[width=0.49\textwidth]{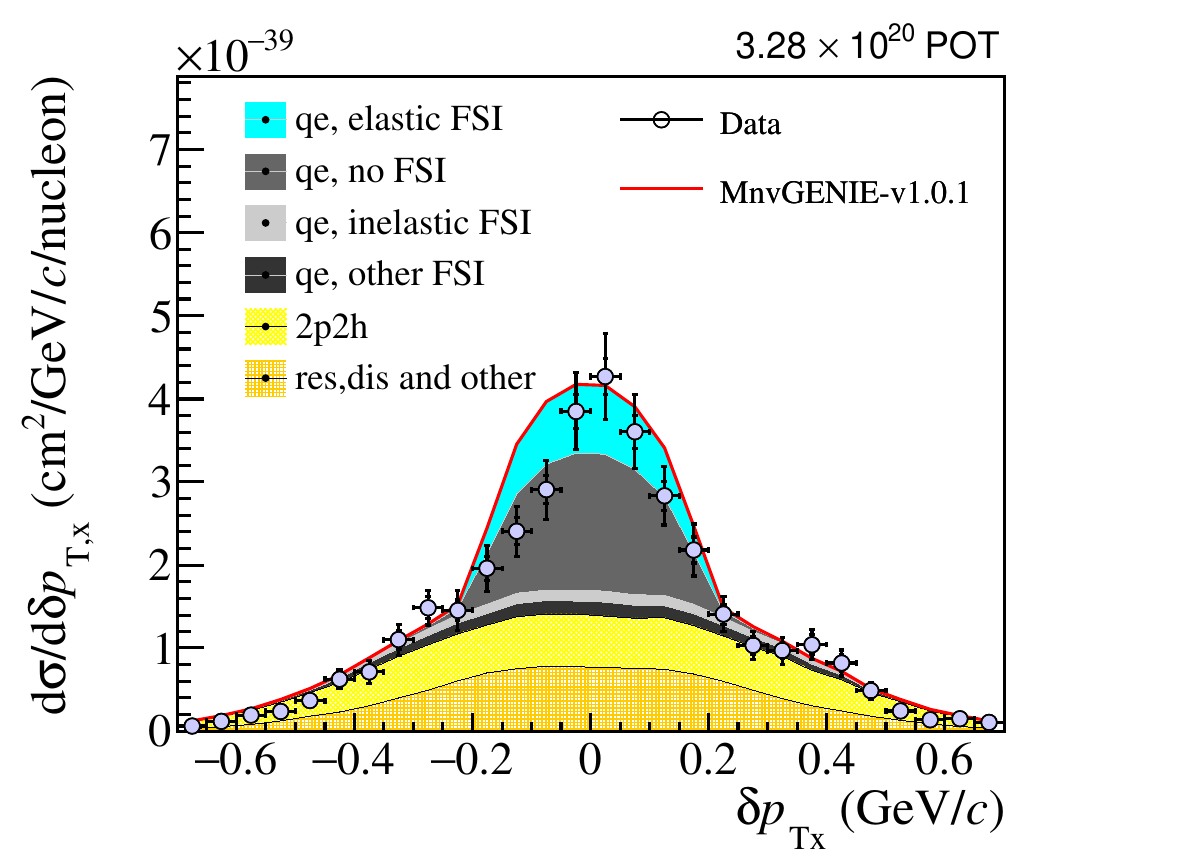}
    \includegraphics[width=0.49\textwidth]{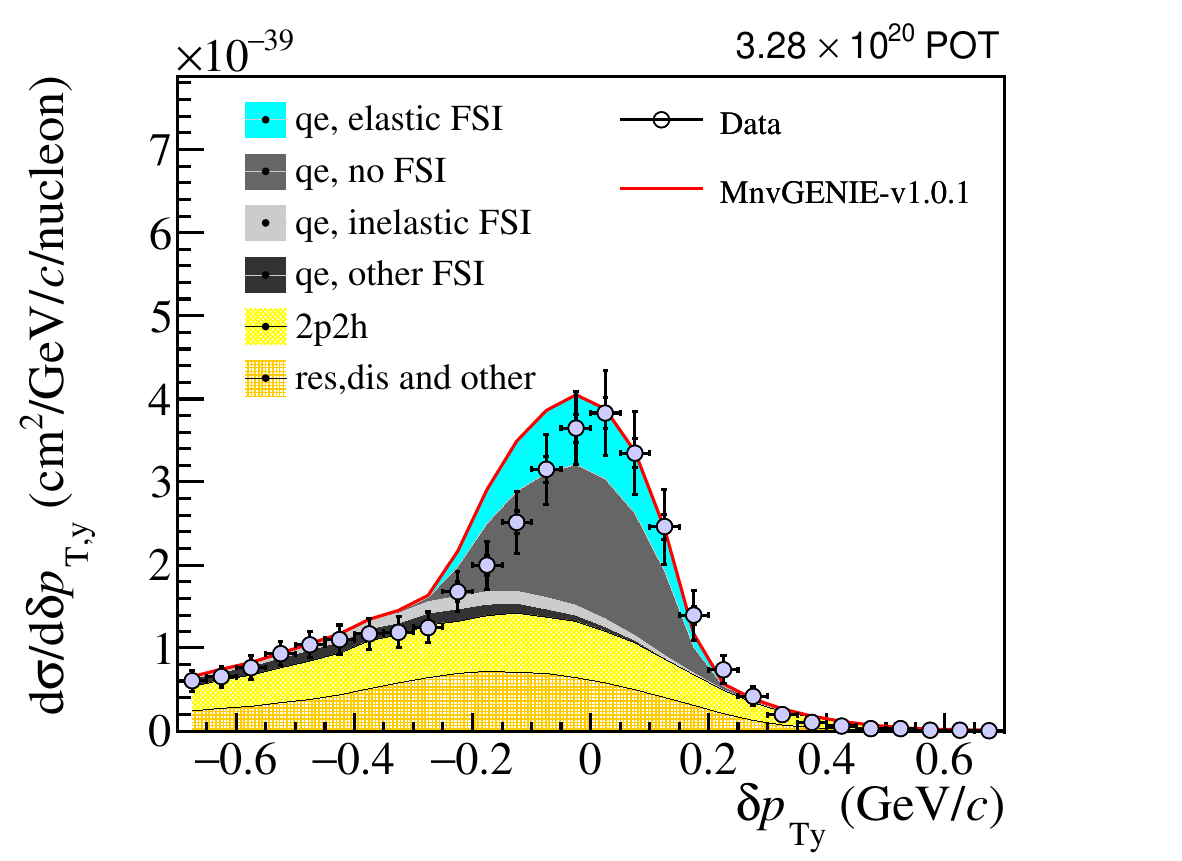}
    \caption{Differential cross sections in $\dptx$ (top) and $\dpty$ (bottom) compared to \mnvgenie{1.0.1}\. The \mnvgenie{1.0.1}\ histogram is separated into the \genie\ defined QE, 2p2h and non-QE event types. The QE region is further separated into the \genie\ FSI experienced by the selected proton before exiting the nucleus. The QE elastic FSI regions displayed in the figures are replaced by the no FSI contributions scaled to 51\%. Note $\dptx$ seems to be slightly asymmetric about the center. The $\dpty$ peak is shifted to the left and has larger width than data.}
    \label{fig:genie}
  \end{figure}
  
  The unfolded cross section results are shown in Fig.~\ref{fig:genie}. The $\dptx$ and $\dpty$ cross sections are in the top and bottom panels respectively. There are significant non-QE contributions for both distributions. Of these about half are due to the tuned 2p2h. For each cross section, the QE distribution is broken down into the generated FSI modes. Here, the \genie\ no FSI means the final state nucleon exited the nucleus without interaction; elastic FSI refers to elastic nucleon-nucleon scattering which typically involves scattering angles less than $10^\circ$; and inelastic FSI refers to events with knockout of one or more additional nucleons. The other FSI category includes channels such as charge exchange multi-nucleon knockout, and pion production/absorption during nucleon transport. Appendix~\ref{app:elasticFSI} describes an error in \genie\!'s implementation of elastic FSI, the fix we implemented, and the effect of the fix on the predictions and the analysis. 
  
\subsection{Distribution in $\dptx$}
  
  \begin{figure}[tbp]
     \centering
     \includegraphics[width=.48\textwidth]{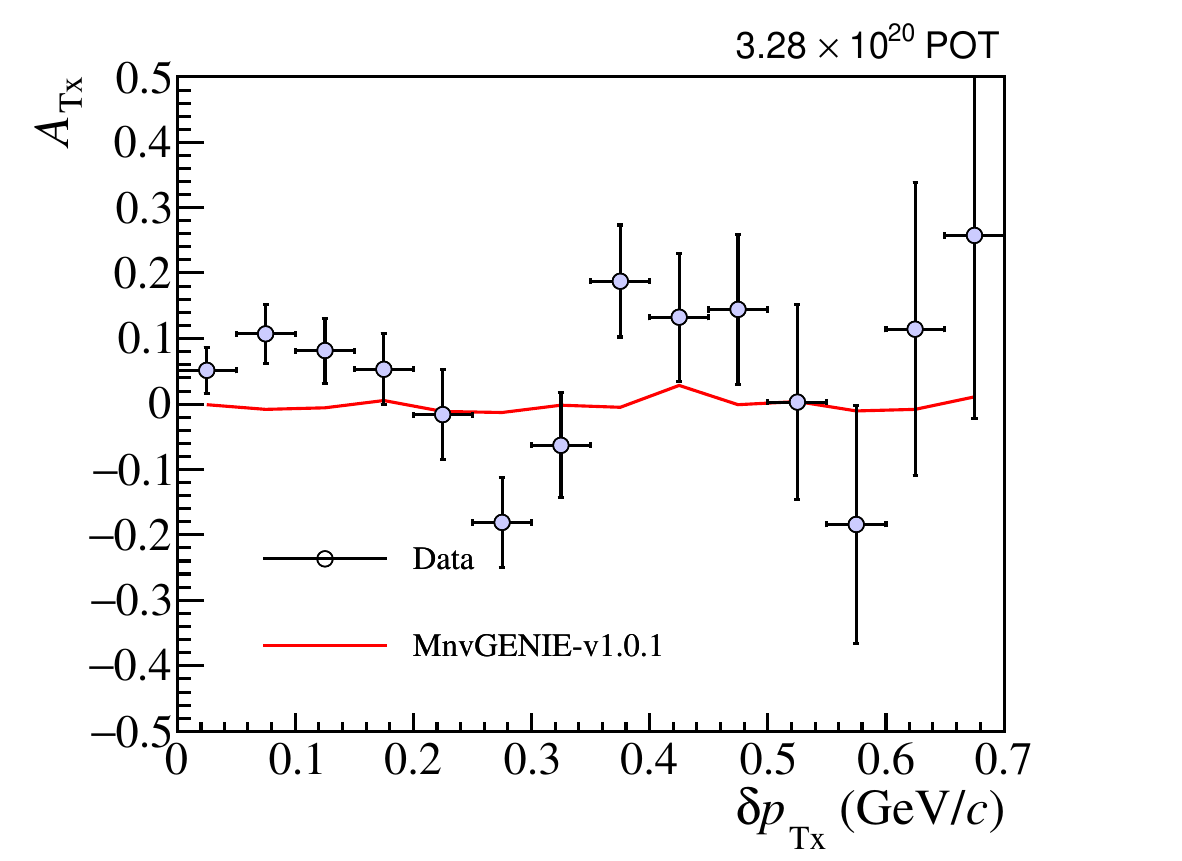}
     \caption{The bin-by-bin asymmetry in the differential cross sections between  $\pm|\dptx|$ bins (Eq.~\ref{eq:atx}). Data is compared to \mnvgenie{1.0.1}, which is representative of the other MC generators used in this study and exhibits no asymmetry.}
     \label{fig:dptxAsym}
 \end{figure}
  
    The measured differential cross-sections  in $\dptx$ and $\dpty$  exhibit a QE peak near $0$. If the interaction occurred on a free nucleon, then we would expect a delta function at $0$ because the muon and proton final states must balance. The width of the QE peak mostly results from Fermi motion. 
   
 The measured cross-section in $\dptx$ in the peak region is wider in the data than in the reference model.  While our correction to simulation of elastic FSI does not precisely reproduce a ``fixed'' elastic FSI, the width of the predicted no FSI contribution itself  is  larger than the data. If we assume no significant deviation in the non-QE distributions, then the discrepancy could imply an overestimation of the Carbon Fermi momentum, or a reduction in the total fraction of the no FSI contribution, or both.
 
 Besides the width discrepancy, the data distribution in $\dptx$ visually leans towards the positive side. 

  To measure the significance of the asymmetry, we define the bin-by-bin asymmetry between the positive and negative sides of the differential cross section in $\dptx$ as:
  \begin{align}
    \atx(|\dptx|) =\frac{\sigma_{+} - \sigma_{-}}{\sigma_{+} + \sigma_{-}}\label{eq:atx},
  \end{align}
  where $\sigma_{\pm}$ is the cross section at either $\pm|\dptx|$ bin. The resulting distribution is reported in Fig.~\ref{fig:dptxAsym}, where observation of bin-by-bin asymmetries in the data and their significances in different ranges of $\dptx$ are reported in Table~\ref{tab:chi2xasym}. None of the generators used in this study reproduce the asymmetric feature, where \mnvgenie{1.0.1}\ is shown as an example. \\
 
  \begin{table}[tbp]
  \caption{$\chi^2$ of asymmetries ($\atx$) against no asymmetry case for regions of $\dptx$ distributions calculated with the covariance matrix.} 
  \label{tab:chi2xasym}
  \centering
  \begin{tabular}{l  l}
  \hline
    $\dptx$ Range ($\gev$)\;\;\;  & $\chi^2$/ndf\;\;\;      \\\hline
    $0.00\sim 0.40$         & $19.9/8$          \\
    $0.40\sim 0.70$         & $4.95/7$          \\
    $0.00\sim 0.70$         & $21.6/15$         \\\hline
  \end{tabular}
  \end{table}
  
  The total asymmetry is defined as:
  \begin{align}
   \alr =\frac{N_{-} - N_{+}}{N_{-} + N_{+}}
    \label{eq:alr},
  \end{align}
  with $N_{-/+}$ being the integrated cross sections on the left/right side of the neutrino-muon plane.  The result is 
  \begin{equation}
      \alr=-0.05\pm0.02,
  \label{eqn:alr_result}
  \end{equation} 
  where the uncertainty is calculated from the covariance matrix in the Supplemental Material. 

  Such an asymmetry has been suggested to result from the pion absorption contributions to the signal~\cite{Cai:2019jzk}. Measurements of single-pion production at low energy in deuterium ~\cite{PhysRevD.25.1161} and single-$\pi^0$ production by \minerva~\cite{PhysRevD.96.072003} have seen positive pion asymmetries about the neutrino-muon plane. The  correlated proton angular distributions in this measurement, from baryon resonance production with an unobserved absorbed pion,   could  exhibit an opposite asymmetry. \\

\subsection{Distribution in $\dpty$}

  Unlike in the $\dptx$ distribution, we observe a non-QE tail towards the negative $\dpty$ values. Inelastic events such as 2p2h, resonance and DIS are inefficient at transferring the lepton momentum to the final state nucleons, since multiple initial states particles are often involved. Therefore the protons tagged in the non-QE events will in general have less momenta then the muons, and are shifted to the left.
  
\begin{figure*}[tbp]
    \includegraphics[width=0.49\textwidth]{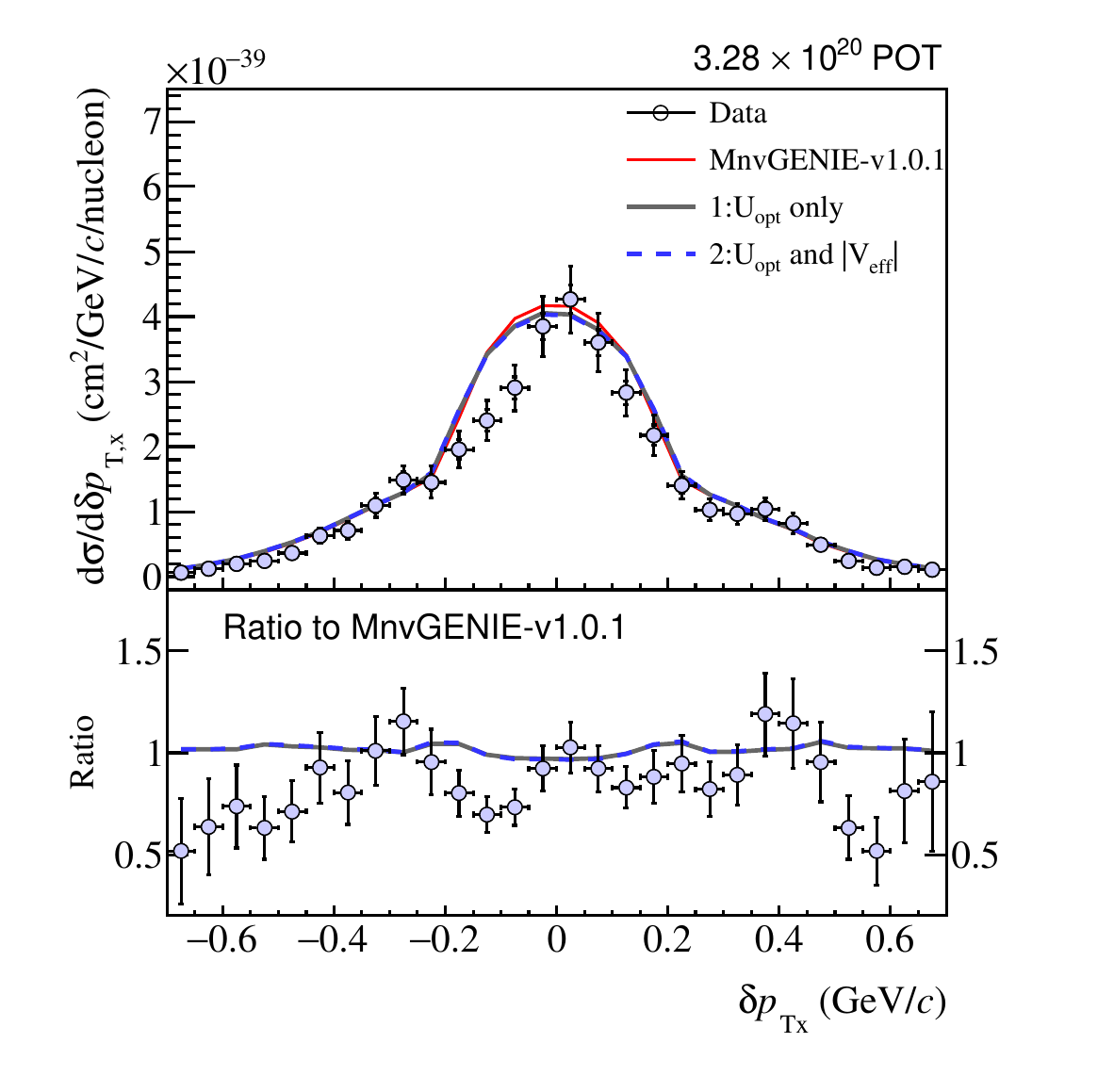}
    \includegraphics[width=0.49\textwidth]{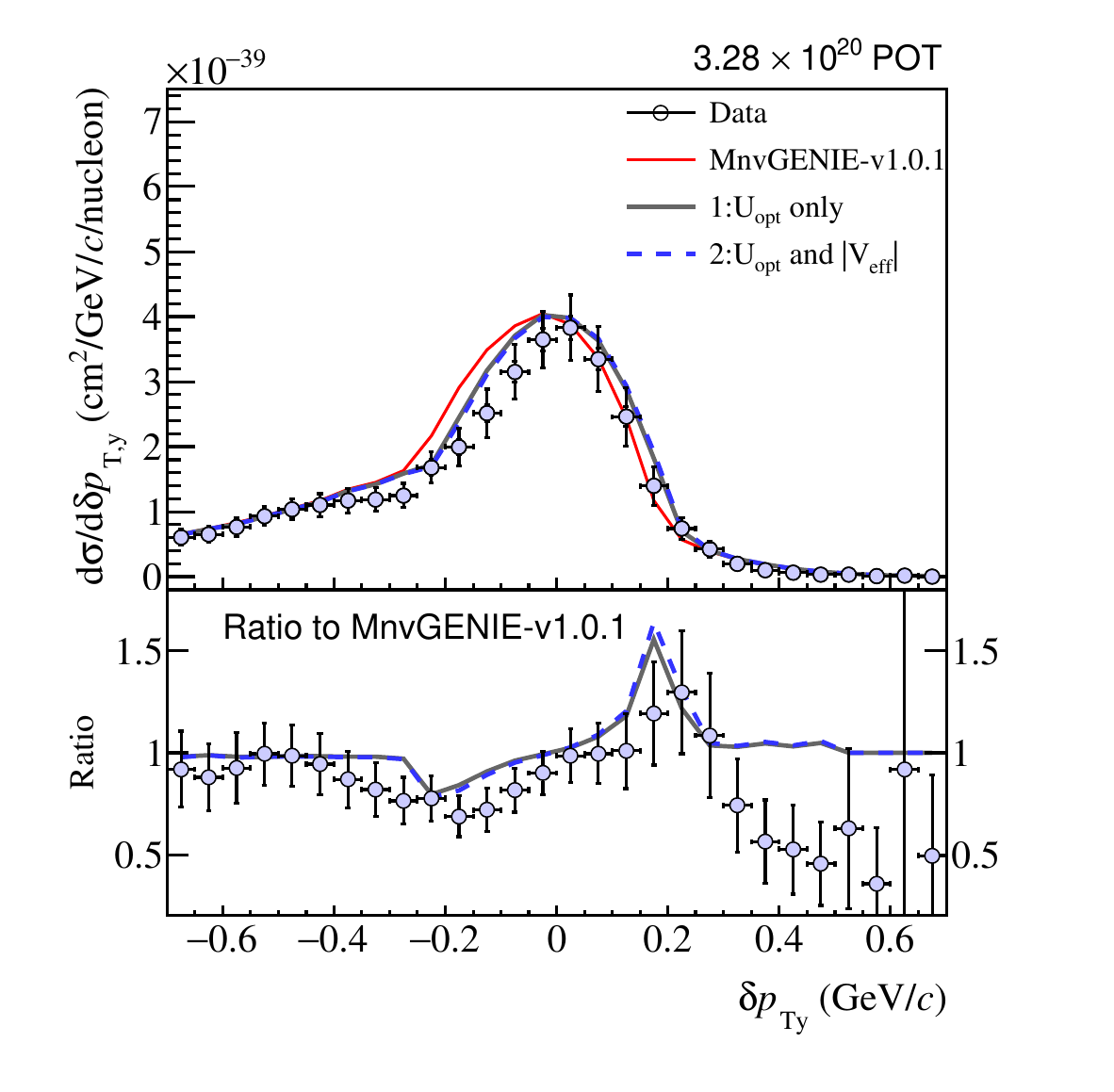}
    \caption{Differential cross sections in $(\dptx,\dpty)$ compared with \mnvgenie{1.0.1}\ \intE\ corrections, defined in Table~\ref{tab:GenieCorrectionModes}. The corrections minimally affect $\dptx$, while bringing the $\dpty$ peak region into closer agreement with data. Note the similar trends in $\dpty$ ratios between the corrections and data.}
    \label{fig:genieCorr}
  \end{figure*}

  The two sets of corrections proposed in Table~\ref{tab:GenieCorrectionModes} are made to the final states muons and protons in \mnvgenie{1.0.1}' CCQE contribution in the MC sample.
 The effect of correction 1, with $\ufsi$ only, and correction 2, with both $\ufsi$ and $\left|\veff\right|$ corrections, are shown in Fig.~\ref{fig:genieCorr}. The effects of $\ufsi$ is on the order of $2\mev$ as it mainly affects nucleons at low kinetic energies. 
  
   Almost all of the shift comes from adding the Moniz interaction energy for Carbon ($\Delta^\text{C}_\genie$) back to the final state proton, and removing the average excitation energy ($\langle E_x\rangle$) from the muon. These corrections alone shift the $\dpty$ peak $34.2\mevc$ to the right. Application of the optical potential partly cancels the shift, resulting in a net shift of $29.4\mevc$.  However, the addition of the Coulomb effect shifts the peak back, nearly canceling the effect of the optical potential, for a net shift including both effects of $33.9\mevc$.

  The ratios, in the lower panels of Fig.~\ref{fig:genieCorr}, of the corrected models and the data to \mnvgenie{1.0.1} show the same upward-going trend in the QE peak region between $\left| \dpty \right| < 0.2\gev$. 
  This trend is characteristic of a peak shift, and the similarities lend confidence to the validity of the theoretically motivated corrections. 
  
\begin{figure*}[tbp]
    \includegraphics[width=0.49\textwidth]{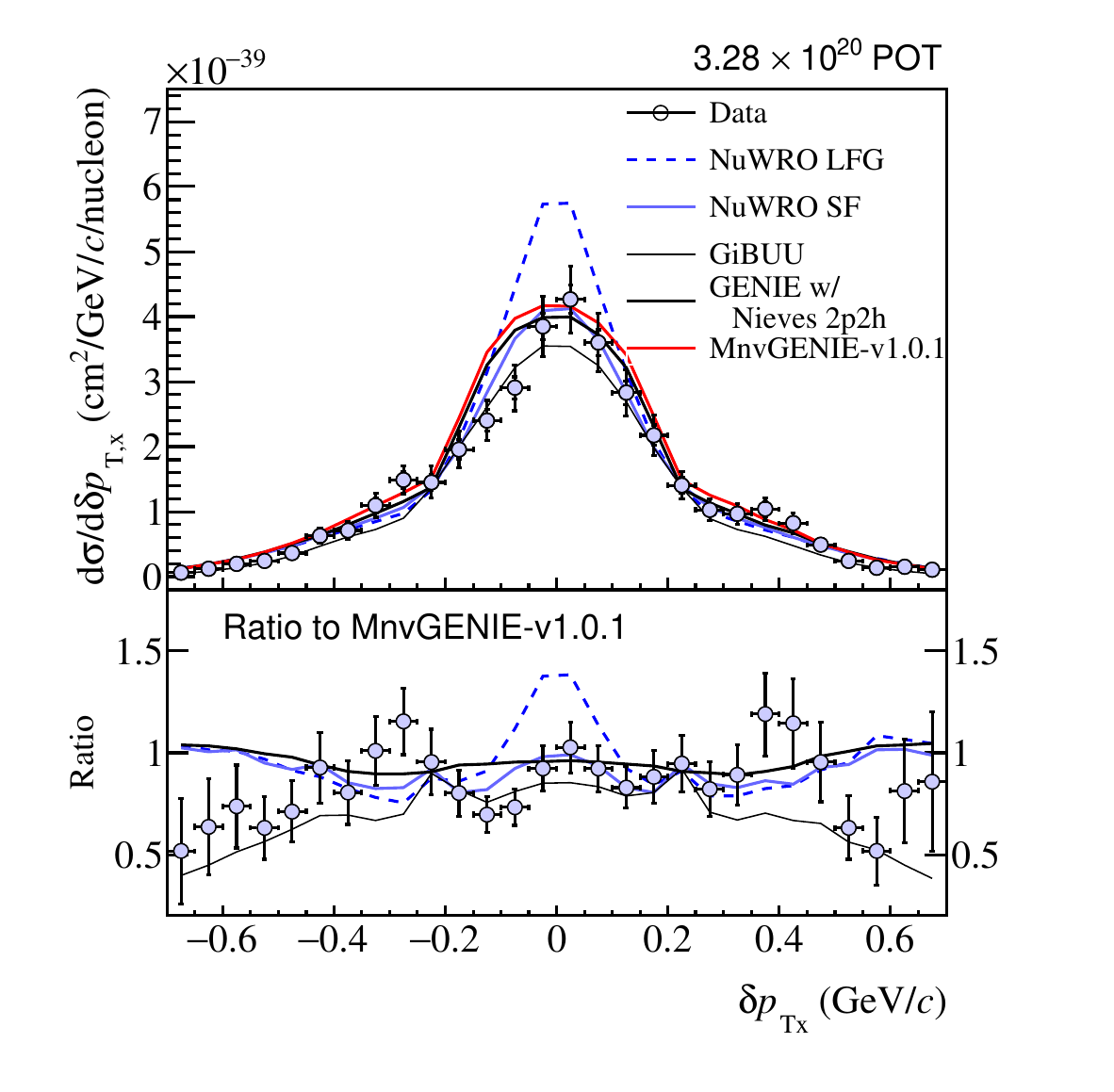}\hspace{-.3cm}
    \includegraphics[width=0.49\textwidth]{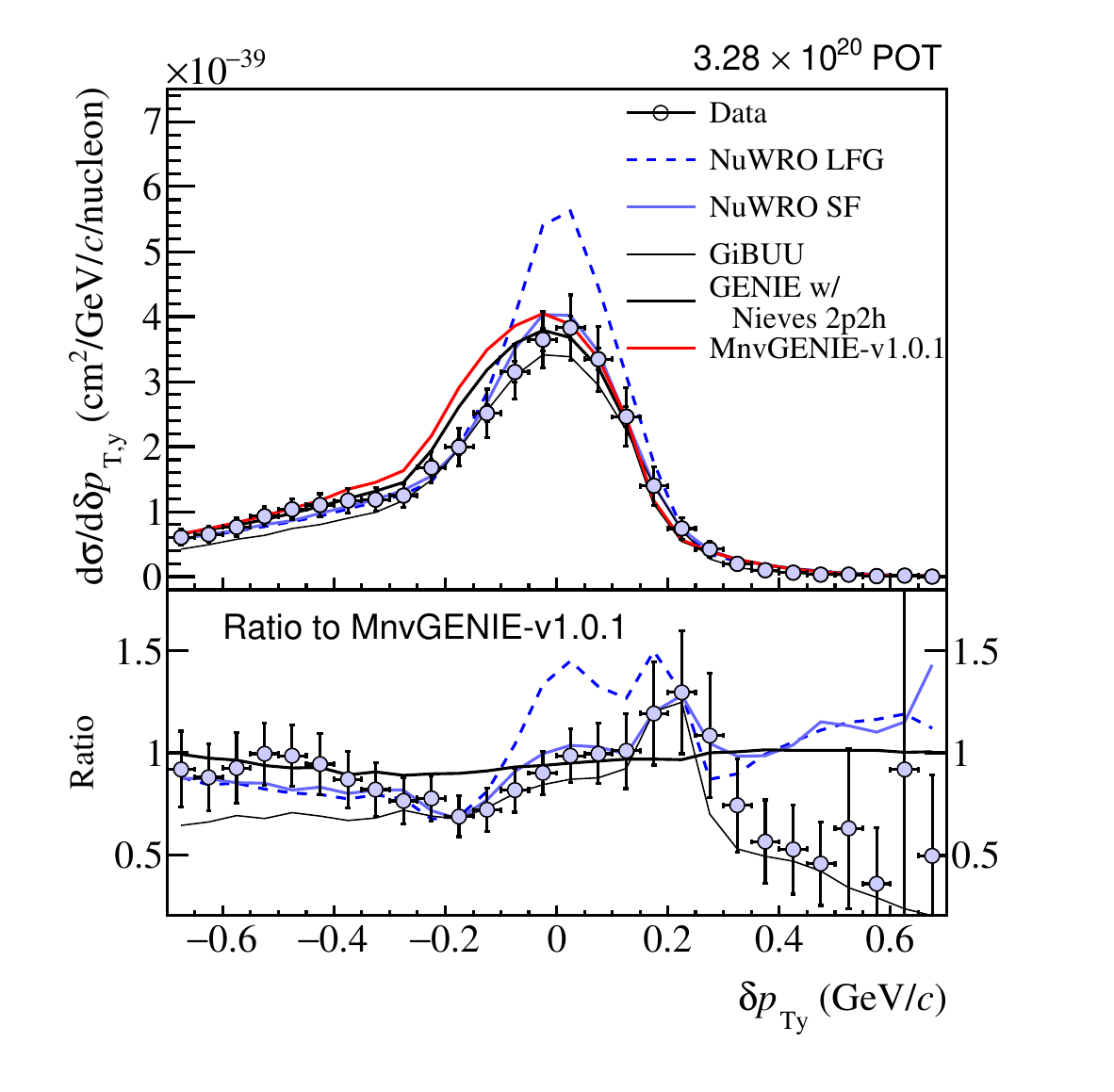}
      \caption{Model comparisons for \nuwro~LF, \nuwro~SF,\gibuu, default \genie\ and \mnvgenie{1.0.1}. Both \nuwro~SF and \gibuu observe peak shifts consistent with data. \nuwro~LFG, although more discrepant from the data, describes the narrowness of the data peak well.  }
      \label{fig:nng}
  \end{figure*}

\begin{figure*}[tbp]
    \includegraphics[width=0.49\textwidth]{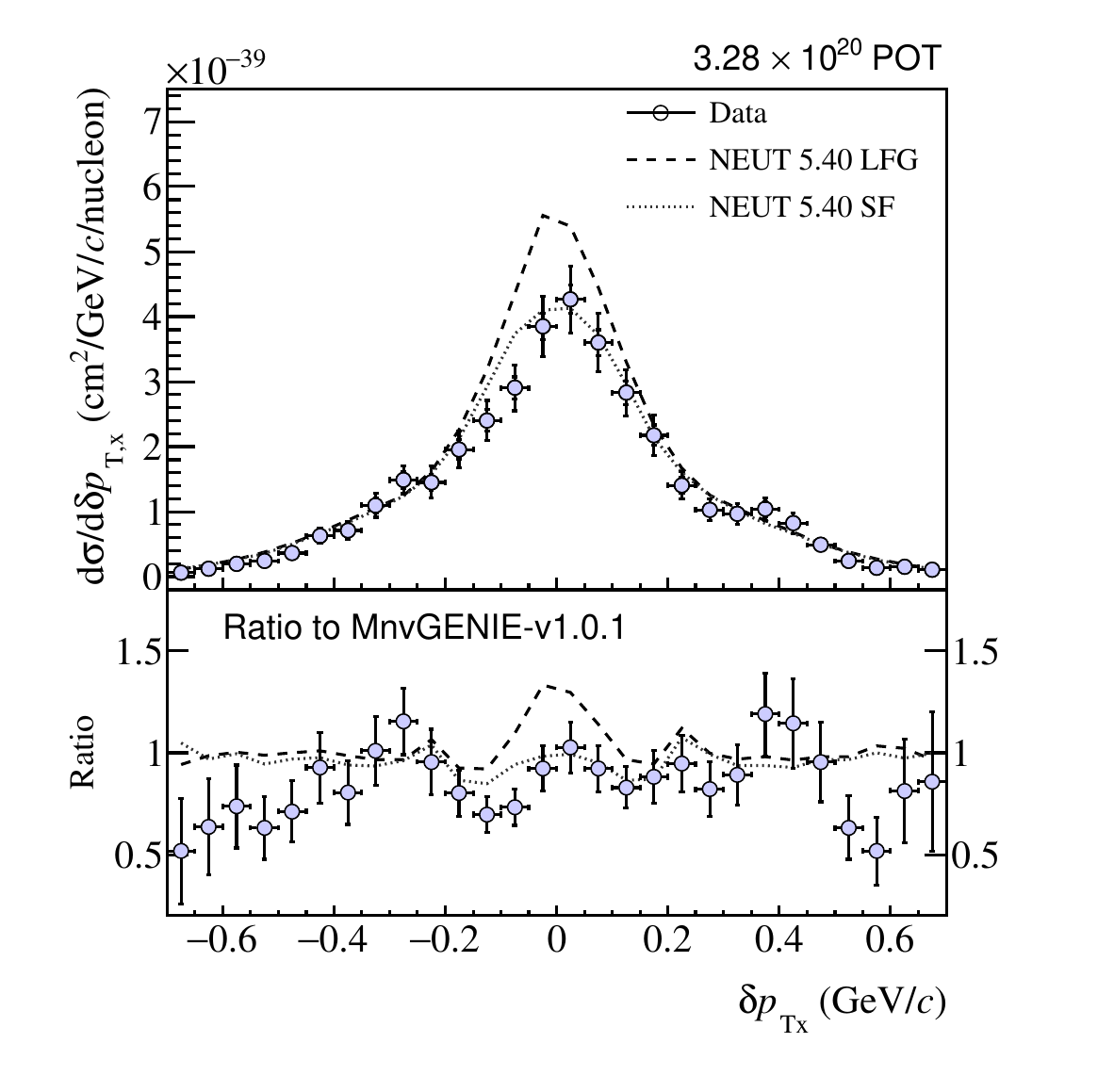}\hspace{-.3cm}
    \includegraphics[width=0.49\textwidth]{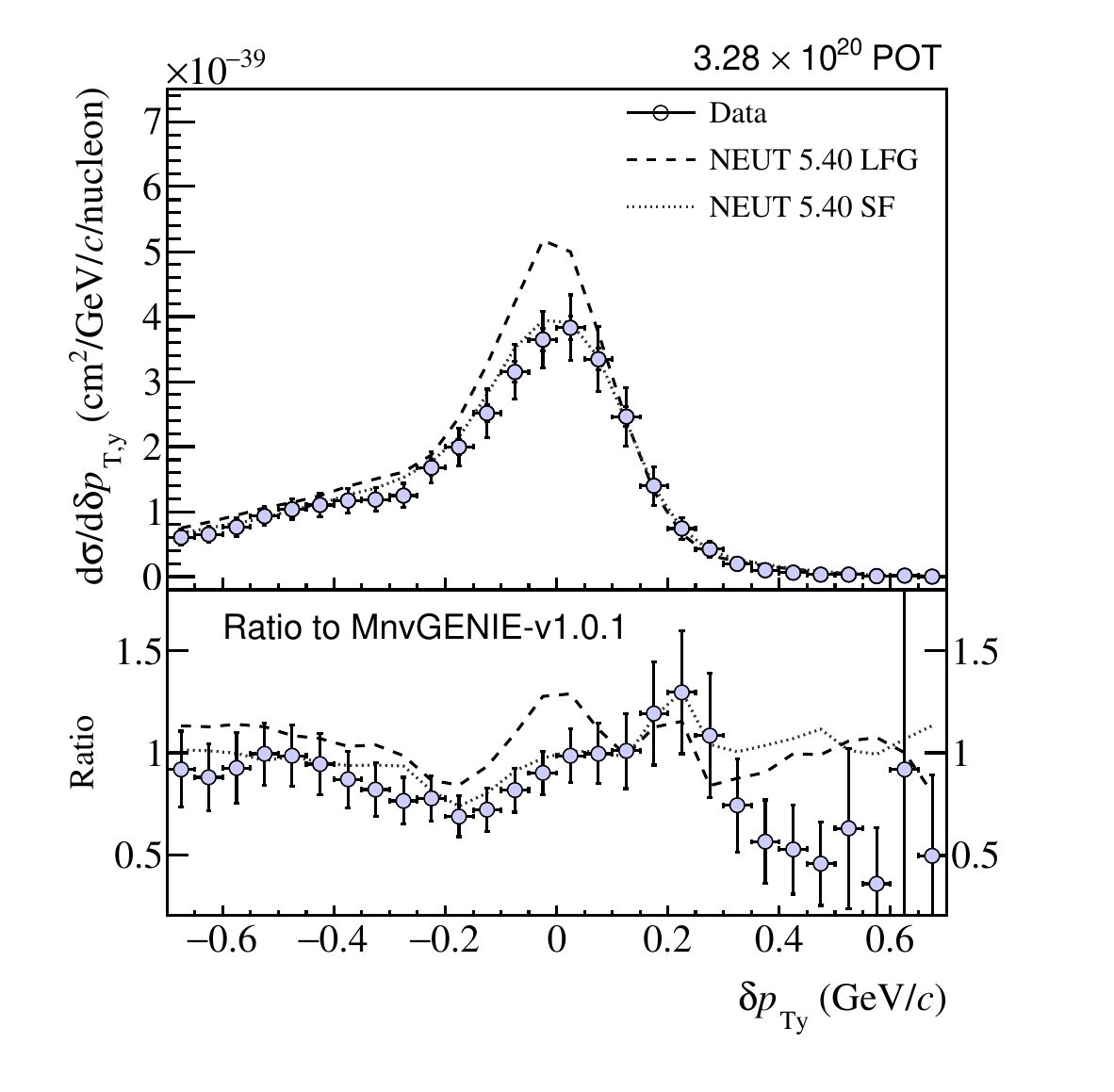}
      \caption{Model comparisons for \neut~SF and \neut~LFG. The $\dpty$ distribution in \neut~SF describes the data peak well, while \neut~LFG over-predicts the left side of the peak, leading to a wider peak similar to \genie. }
      \label{fig:neut}
  \end{figure*}

  Figure~\ref{fig:nng} and Fig.~\ref{fig:neut} compares \nuwro\ Local Fermi Gas (LFG), \nuwro\ Spectral Function (SF), \gibuu, the nominal \genie, \mnvgenie{1.0.1}, as well as \neut~SF and LFG, distributions normalized to data cross sections. In terms of $\dpty$, the nominal \genie\ with Nieves 2p2h does not depart much from the overall peak offset seen in \mnvgenie{1.0.1}, the ratio between which is nearly flat.  The modifications to the 2p2h fraction, the non-resonant pion reweighting and RPA introduced by the \minerva\ tune have little effect on the position of the peak, since their effects are nearly constant at the QE peak.
  Data to \mnvgenie{1.0.1} ratio, and in fact the ratios of all other models to \mnvgenie{1.0.1}, except \nuwro\ LFG follow very similar trends. The \nuwro~SF and \gibuu~models both have better agreements with data while \nuwro~LFG has overall disagreement in cross section. 
  
  The \nuwro\ models include nuclear effects such as Pauli blocking and the Coulomb potential. The \nuwro\ SF model, in particular, includes an effective potential simulating the optical potential. The effective potential is validated against electron scattering data on targets including $^{16}O$~\cite{Ankowski:2007uy}, a nucleus similar to $^{12}C$~\cite{Bodek:2018lmc}. The \nuwro\ LFG has larger disagreement with the data. However, the average Fermi motion of the typical LFG models produces a narrower width in the QE peak than that of the Fermi momentum in regular Fermi gas models.  This produces a narrower peak more suggestive of the data.
  
  The \neut\ SF describes the QE peak location well, while the LFG shifts the peak location by more than $1\sigma$. In fact \neut\ SF describes both $\dptx$ and $\dpty$ very well near the peak regions.Unlike the \nuwro\ variant, \neut~LFG predicts wider QE width in $\dpty$ while at the same time produces width in $\dptx$ comparable to that of the data.
  
  \gibuu\ models the initial state nucleons with a local Thomas-Fermi approximation, and the nucleons are bound in a mean-field potential, where Pauli blocking is naturally simulated.  The final state particles propagating through the nuclear medium are subject to a scalar potential that usually depend on both the nucleon momentum and nuclear density~\cite{Leitner:2010kp}. These features of \gibuu\ do not contribute to an especially superior description of the QE peak. Unrelated to the description of the peak, the tail distributions of the \stki\ quantities are sensitive especially to the 2p2h component and pion production followed by pion absorption with proton knockout. With a lower proton threshold than this analysis, it could include significant amounts of QE events followed by FSI. \gibuu\ seems to be quite adept at predicting three of the four tails of these signal distributions, while the other generators systematically overestimate these regions.

  We investigate the agreement of the $\dpty$ measurement with model predictions using a weighted average, $\dptyavg$, defined as
  \begin{align}
     \dptyavg &= \frac{\sum_i \sigma_i{\dpty}_i}{\sum_j \sigma_j}, \\
     V &= \frac{\sum_{i,j} {\dpty}_i C_{ij} {\dpty}_j}{(\sum_k \sigma_k)^2  }, 
  \end{align}
  where $\sigma_i$ and ${\dpty}_i$ are the cross section and position in the $i$-th $\dpty$ bin, $i,j$ span over the summed range. The calculation of the variance $V$ takes into account the covariance matrix $C_{ij}$, which contains the correlated errors between the $i$th and $j$th bins. The covariance matrices for $\dptx$, $\dpty$ and the variables reported in Ref~\cite{Lu:2018stk} are available as digital data release.

\begin{figure}[tbp]
      \centering
      \includegraphics[width=.48\textwidth]{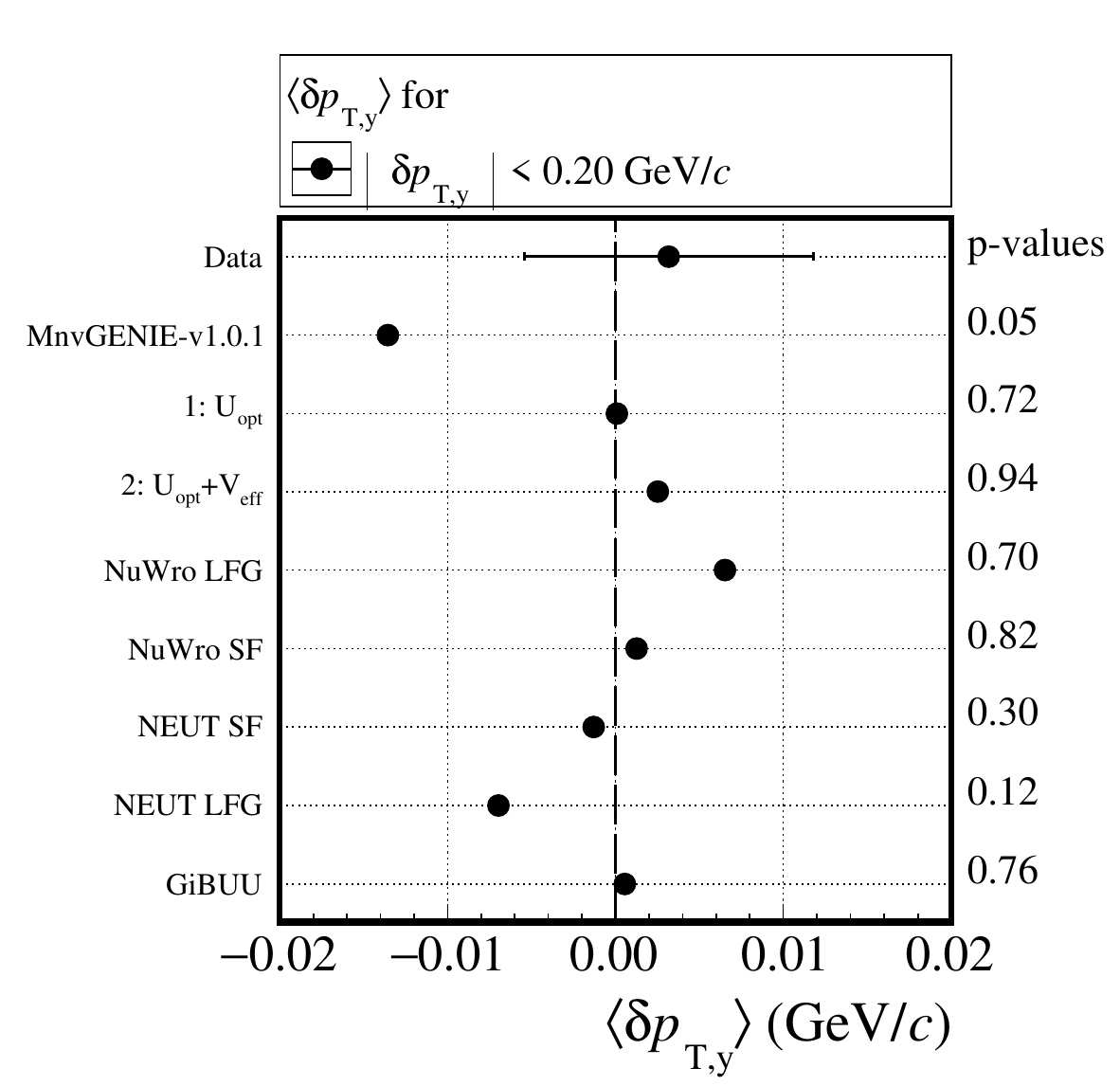}
      \caption{$\left<\dpty\right>$ calculated from the differential cross section within $|\dpty|<0.20\gevc$. The p-value is the probability, assuming normal distribution, that the observed result would have been produced by change from this model.}
      \label{fig:dpTyAvg}
  \end{figure}

  The computation of $\dptyavg$ is sensitive to the range selected due to the underlying non-QE contribution. The $\range{-0.20}{0.20}\gevc$ momentum range is chosen because it is dominated by the QE events. The results are summarized in Fig.~\ref{fig:dpTyAvg}.
  
  
  For each model, a p-value is calculated under the assumption of normally distributed uncertainties on the data. The average peak positions of \mnvgenie{1.0.1}\ lie outside $1\sigma$ uncertainty range of the data. Measurable shifts to larger $\dptyavg$ are observed when \intE\ corrections are applied. The shifts are on the order of $15$ to $20 \mevc$, consistent with corrections made to the underlying model. The measurements disfavor the default \genie\ \removalE\ implementation, but does not distinguish between the nuclear potential corrections. Among the models \nuwro\ SF, \neut\ SF and \gibuu\ models are comparable to the data average, while \nuwro\ and \neut\ LFGs have larger disagreement with the data. Between them, \neut\ LFG peaks outside the measurement uncertainties.

\begin{table*}[!tbh]
     \centering
      \caption{$\dpty$: $\chi^2$ comparisons, POT normalized.}
      \label{tab:chi2y}
      \begin{tabular}{l|c|>{\bfseries}c>{\bfseries}c|c||c|c}\hhline{|=======|}
        POT Normalized   & -0.2 $\sim$ -0.1 GeV  & -0.1 $\sim$ 0.0 GeV   & 0.0 $\sim$ 0.1 GeV  & 0.1 $\sim$ 0.2 GeV  & -0.2 $\sim$ 0.2 GeV & -0.7$\sim$ 0.7\\ \hline
        \genie\ Nominal           & 41.1/2  & 19.0/2  & 0.743/2   & 13.5/2  & 52.9/8    & 69.5/28\\ \hhline{|=|=|==|=|=|=|} 
        
        \mnvgenie{1.0.1}          & 89.0/2  & 38.9/2  & 0.184/2   & 13.7/2  & 100/8   & 72.5/28\\\hline
        1: $\ufsi$ only          & 32.4/2  & 22.5/2  & 2.73/2   & 18.8/2  & 38.2/8  & 111/28    \\ \hline 
        2: $\ufsi$ and $\veff$   & 27.7/2  & 19.6/2  & 3.71/2   & 30.2/2  & 45.6/8  & 111/28      \\ \hhline{|=|=|==|=|=|=|} 
        \nuwro\ LFG               & 25.1/2  & 159/2   & 130/2   & 15.5/2  & 50.7/8 & 131/28 \\ \hline 
        \nuwro\ SF                & 10.6/2  & 8.87/2   & 1.46/2   & 0.296/2   & 6.66/8  &60.0/28\\  \hhline{|=|=|==|=|=|=|} 
        \neut\ 5.40 LFG           & 43.6/2  & 113/2   & 82.6/2  & 0.842/2   & 52.6/8 & 75.9/28\\ \hline 
        \neut\ 5.40 SF            & 7.31/2   & 9.03/2   & 0.397/2   & 0.302/2   & 4.41/8 & 54.3/28\\ \hhline{|=|=|==|=|=|=|} 
        \gibuu\                   & 1.50/2   & 3.81/2   & 6.85/2   & 6.04/2   & 7.70/8 & 45.0/28\\  \hhline{|=|=|==|=|=|=|} 
      \end{tabular}
\end{table*}

  Next, we calculate $\chi^2$ distributions in four consecutive, disjoint $\dpty$ ranges dominated by QE interactions to illustrate the mismodelling in the \mnvgenie{1.0.1}\ simulations.
  Table.~\ref{tab:chi2y} summarizes the results.  The $\chi^2$ in $\dpty$ for \mnvgenie{1.0.1}\ is not symmetric about 0, where the falling side $\range{0}{0.2}\gevc$, with $\chi^2 = 13.7$, is in much better agreement with the data than the rising side $\range{-0.2}{0}\gevc$ with $\chi^2 = 89.0$. 
  
  The corrections for \mnvgenie{1.0.1}\ reduce the model asymmetry, bringing the $\chi^2$ at the left edge from $89.0$ to the order of $30$. The $\chi^2$ for the right edge increases from $13.7$ to $18.8$ and $30.2$ between corrections 1 and 2. The total $\chi^2$ between $\range{-0.2}{0.2}\gevc$ is reduced by more than $50\%$ after the corrections are applied. The overall $\chi^2$s for \mnvgenie{1.0.1} is $72.5$, while both its corrections are $111$ for 28 degrees of freedom.
  
  Other Fermi gas-based models, such as \nuwro\ LFG and \neut\ LFG, in general have better $\chi^2$ than the \genie\ variations. The \nuwro\ LFG $\chi^2$ in the edges are more consistent with each other, at $25.1$ and $15.5$ respectively. The \neut\ LFG, on the other hand, seems to suffer from model asymmetry similar to \mnvgenie{1.0.1}, but the cause might be due to a systematic excess in cross section predicted in the negative tail of the $\dpty$ distribution, as shown in Fig.~\ref{fig:neut}. The tail is dominated by non-QE interactions. 
  In contrast, spectral function models and \gibuu\ predict $\dpty$ very well.

  \begin{table*}[!tbh]
     \centering
      \caption{$\dptx$: $\chi^2$ comparisons, POT normalized.}
      \label{tab:chi2x}
      \begin{tabular}{l|c|>{\bfseries}c>{\bfseries}c|c||c|c}\hhline{|=======|}
        POT Normalized   & -0.2 $\sim$ -0.1 GeV  & -0.1 $\sim$ 0.0 GeV   & 0.0 $\sim$ 0.1 GeV  & 0.1 $\sim$ 0.2 GeV  & -0.2 $\sim$ 0.2 GeV & -0.7 $\sim$ 0.7 GeV \\ \hline
        \genie\ Nominal           & 26.0/2  & 31.6/2  & 3.40/2   & 4.03/2   & 26.4/8    & 69.5/28   \\ \hhline{|=|=|==|=|=|=|} 
        \mnvgenie{1.0.1}          & 38.6/2  & 40.4/2  & 4.00/2   & 9.11/2   & 34.5/8    & 67.2/28\\\hline
        01: $\ufsi$ only          & 36.3/2  & 35.2/2  & 4.02/2   & 9.40/2   & 35.0/8    & 67.4/28     \\ \hline 
        02: $\ufsi$ and $\veff$   & 36.2/2  & 34.4/2  & 4.03/2   & 9.55/2   & 35.2/8    & 67.8/28    \\ \hhline{|=|=|==|=|=|=|} 
        \nuwro\ LFG               & 22.2/2  & 85.5/2  & 31.4/2  & 4.72/2   & 58.7/8     & 132/28 \\ \hline 
        \nuwro\ SF                & 8.79/2   & 20.1/2  & 0.831/2   & 1.48/2   & 16.6/8  & 63/28 \\  \hhline{|=|=|==|=|=|=|} 
        \neut\ 5.40 LFG           & 21.4/2  & 73.3/2  & 19.0/2  & 5.82/2   & 43.5/8     & 85.2/28 \\ \hline 
        \neut\ 5.40 SF            & 10.2/2  & 24.8/2  & 1.36/2   & 0.632/2   & 17.5/8   & 58.3/28     \\ \hhline{|=|=|==|=|=|=|} 
        \gibuu\                   & 1.69/2   & 11.7/2  & 7.69/2   & 1.27/2   & 11.9/8   & 40.6/28 \\  \hhline{|=|=|==|=|=|=|} 
      \end{tabular}
  \end{table*}
  We also show the $\chi^2$ distributions for $\dptx$ in Table~\ref{tab:chi2x}. Across all of the models we observe bias in the $\chi^2$ as a result of the asymmetry in data that we previously characterized by the measurement of $\alr$ in Eqn.~\ref{eqn:alr_result}.

  \section{Summary and outlook\label{sec:conclusion}}
  
  The variables $\dptx$ and $\dpty$ are measured on the CH target in \minerva. We expect $\dptx$ to be sensitive to the Fermi momentum in QE and there is tension between data and MC. The data is narrower than the \genie\ model, as is true of most models other than a simple Fermi gas. The measurement also shows a statistically marginal proton asymmetry in $\dptx$ of $-0.05\pm 0.02$. This asymmetry, if truly non-zero, might be attributed to pion absorption events included in the signal. No model in current event generators predicts an asymmetry. Future measurements could verify the presence of this asymmetry.
  
  The observable $\dpty$ shows sensitivity to the \intE\ implemented in nuclear models. In particular, the measurement, which is based on \genie, disfavors the default \genie\ implementation of the \intE\ on Carbon. This implementation lacks the excitation energy while subtracts an extra Moniz interaction energy from the final state proton. The average peak positions between \genie\ and data differ by more than 1.5$\sigma$. Approximate corrections accounting for the excitation energy and Moniz interaction energy bring the average peak position within 1$\sigma$ of the data. This measurement is not precise enough to distinguish the more subtle nuclear effects such as the optical potential and the Coulomb potential. To first order more statistics could reduce the overall uncertainties in the distributions. Further improvements in the overall uncertainties need to come from better constrained flux, detector response and signal model, especially in the modelling of pion absorption in the nucleus.
 
  We have compared different Monte Carlo models with respect to $\dptx$ and $\dpty$. The measurements are based on the \mnvgenie{1.0.1}~tune of \genie, which removes the elastic FSI components in \genie\ on top of the \mnvgenie{1}\ base tune. This modification subsequently impacts the \stki\ measurements performed in Ref~\cite{Lu:2018stk}. The elastic FSI is discussed in Appendix~\ref{app:elasticFSI}. 
  The Supplemental Material to this paper contains an update to the \stki\ results presented in Ref~\cite{Lu:2018stk} based on this modification.
  
  Future \minerva\ analysis using the medium energy ~\cite{numi-me} dataset will benefit from higher statistics, which will enable examination of correlations between $\dptx$, $\dpty$ and other variables.  In particular, probing the correlation of the asymmetry in $\dptx$ with other variables may shed light on its origin. Other targets in \minerva\ and future liquid argon experiments could make measurements of $(\dptx,\dpty)$, and the \stki\ variables in general, to test models on other nuclei.

  This work was supported by the Fermi National Accelerator Laboratory under US Department of Energy contract No. DE-AC02-07CH11359 which included the \minerva\ construction project. Construction support was also granted by the United States National Science Foundation under Award PHY-0619727 and by the University of Rochester. Support for participating scientists was provided by NSF and DOE (USA), by CAPES and CNPq (Brazil), by CoNaCyT (Mexico), by Proyecto Basal FB 0821, CONICYT PIA ACT1413, Fondecyt 3170845 and 11130133 (Chile), by DGI-PUCP and UDI/VRI-IGI-UNI (Peru), by the Latin American Center for Physics (CLAF), by Science and Technology Facilities Council (UK), and by NCN Opus Grant No. 2016/21/B/ST2/01092 (Poland). We thank the MINOS Collaboration for use of its near detector data. We acknowledge the dedicated work of the Fermilab staff responsible for the operation and maintenance of the beam line and detector and the Fermilab Computing Division for support of data processing.

\bibliography{bibliography}

\appendix
 
\section{Derivations of GENIE corrections}\label{app:genieCorr} 
  The purpose of this correction is to modify the prediction of the \genie event generator for a different value of $\exN$.  In general, this correction could modify both the energy and three-momentum transferred to the nucleus, but have the freedom to pick some quantity which should be conserved event-by-event in this correction.  We choose the magnitude of the three momentum transfer, $q_3$.  Changes to $q_0$, $Q^2$, and angles and energies of the final state muon and proton follow.  Denote the change in $q_0$ to be
  \begin{equation}
  \tau=\exN-|\ufsi[\vec{(k+q_3)}^2]| + |\veffp|,
  \end{equation}
  and let
  \begin{equation}
  \mnprime=M_N-S^N - \frac{\left<\vec{k}^2\right>}{2M_{A-1}^2}.
  \end{equation}

  Then the energy conservation at the vertex in \genie\ is
  \begin{equation}
      \qzero_\genie+\mnprime= \sqrt{(\vec{k}+\vec{q}_3)^2+M_P^2},
      \label{eq:1p1h-genie}
  \end{equation}
  comparing to Eq.(\ref{eq:1p1h}), we obtain 
  \begin{equation}
      \qzero\approx \qzero_\genie + \tau.
  \end{equation}
  The difference in energy transfer manifests on the outgoing muon energy:
  \begin{equation}
      E_\nu - E_\mu = E_\nu - E_\mu^\genie + \tau.
  \end{equation}
  we obtain the energy correction to the \genie\ muon:
  \begin{align}
      E_\mu = E_\mu^\genie - \tau
      \label{eq:Ecorr_lep}
  \end{align}
  The outgoing proton energy in \genie\ is
  \begin{equation}
      E^\genie_P = \sqrt{(\vec{k}+\vec{q}_3)^2+M_P^2} - \dN{nucleus}.
  \end{equation}
  Comparing to the right hand side of Eq.(\ref{eq:1p1h}), we obtain
  \begin{equation}
      E_P = E_P^\genie+\dN{nucleus}-|\ufsi|+|\veffp|.
  \end{equation}

  As noted above, this correction conserves energy, and assumes $|\vec{q}_3|$ is constant. The fractional change to the $Q^2$ of the system is approximately $\Delta Q^2/Q^2 = \tau/M_N$. For our sample, $\tau\approx 10 \mev$, produces a $1\%$ shift in $Q^2$, which causes changes to the hard scattering cross-section $\lesssim1\%$.
  
  We can also evaluate how changes in angle of the muon and the proton that are neglected in the \genie\ correction would affect the prediction. The muon momentum before and after the correction are:
      \begin{align}
        \vec{p}_\mu&=\pmu
        \begin{pmatrix}
         0 \\ \sin\left(\theta\right) \\ \cos\left(\theta\right)
        \end{pmatrix}, \\
        \vec{p}_\mu'&=(\pmu-\tau)
        \begin{pmatrix}
         0 \\ \sin\left(\theta+\delta\right) \\ \cos\left(\theta+\delta\right)
        \end{pmatrix} 
        \label{eq:dq3sq}
    \end{align}
  
  Solving the equation $|\vec{p}_\nu-\vec{p}_\mu|^2=|\vec{p}_\nu-\vec{p}_\mu'|^2$ for $\delta$ to first order in $\tau$, we have:
  \begin{align}
      \delta&\approx \tau\left(\frac{1}{\enu\sin(\theta)} - \frac{1}{\pmu\tan(\theta)}\right). 
  \end{align}
  Note that the effect on the angle could become significant at small $\theta$, but in this region $E_\nu-E_\mu$ becomes small, and in this region the recoiling protons in quasielastic events are also soft  that such they do not enter our sample.  For our events, $\delta\lesssim 1.5\mathrm{mrad}$.

  For an interaction, the $\vec{p}_\mu$ and $\vec{p}_P$ are balanced at the vertex. Changing the $\theta$ must elicit a compensating change in the proton angle to conserve the momentum. In our correction, we neglect the small changes of angles above.  This introduces a very small error in the calculation of $\dpty$ for events that pass our selections, particularly the proton momentum cut. This error decreases with $Q^2$ and is at most $0.25\mev$ for $Q^2=0.2\gev^2$. Therefore the simplifying assumption in our modification of \genie\ that the muon and proton angles do not change is justified.
  
\section{GENIE elastic FSI simulation}\label{app:elasticFSI}
This section will discuss the elastic FSI prediction and the fixes to it in more detail. 
  \begin{figure}[tbp]
      \centering
      \includegraphics[width=.4\textwidth]{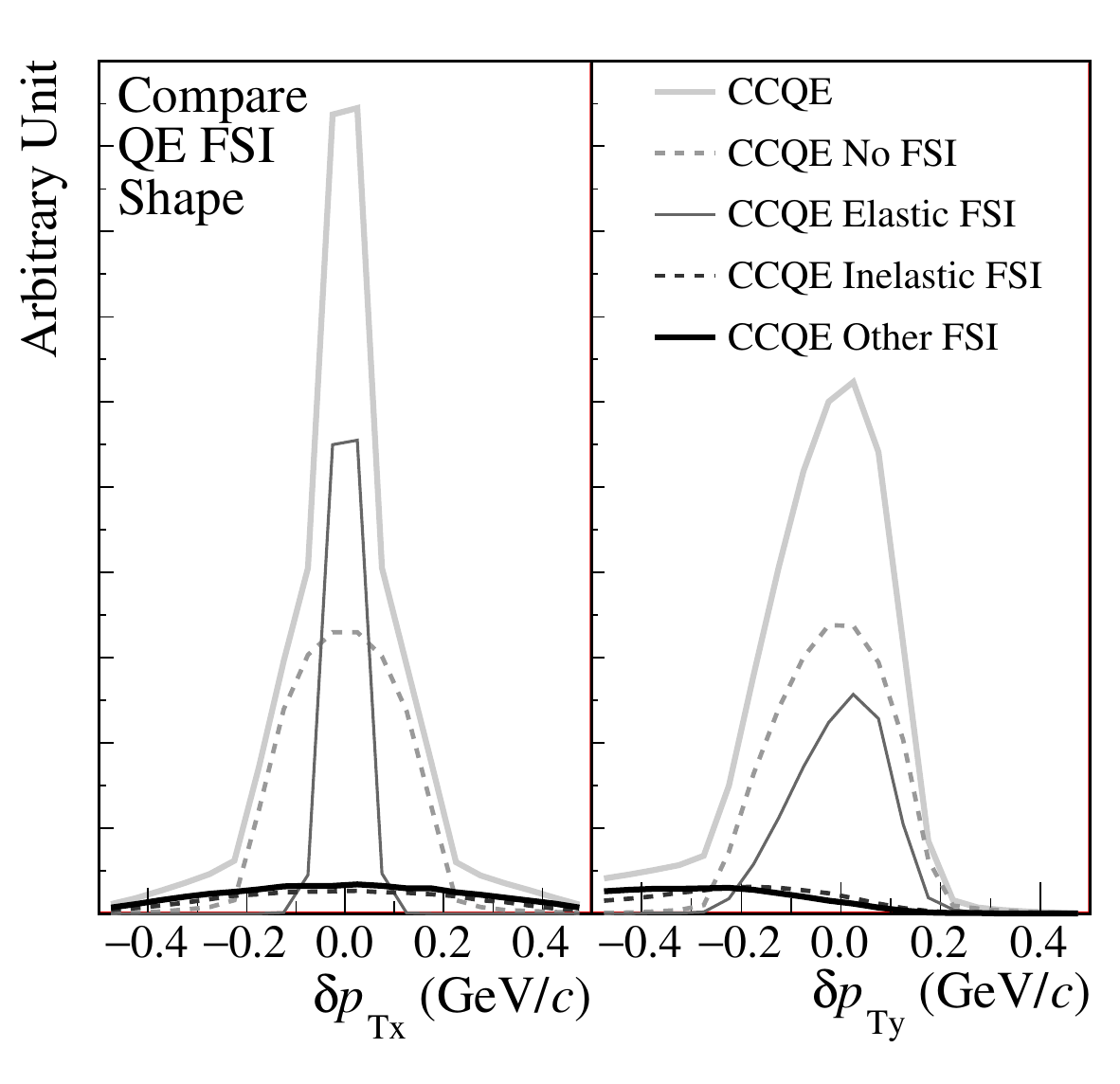}
      \caption{\genie\ FSI modes breakdown for CCQE events. The non-interacting fraction is symmetric and preserves the Bodek-Ritchie Distribution while the \genie\ elastic FSI appears accelerated with respect to the transverse momentum transfer $q_T$. }
      \label{fig:fsi}
  \end{figure}
  
   \begin{figure*}[tbp]
       \centering
       \includegraphics[width=0.3\textwidth]{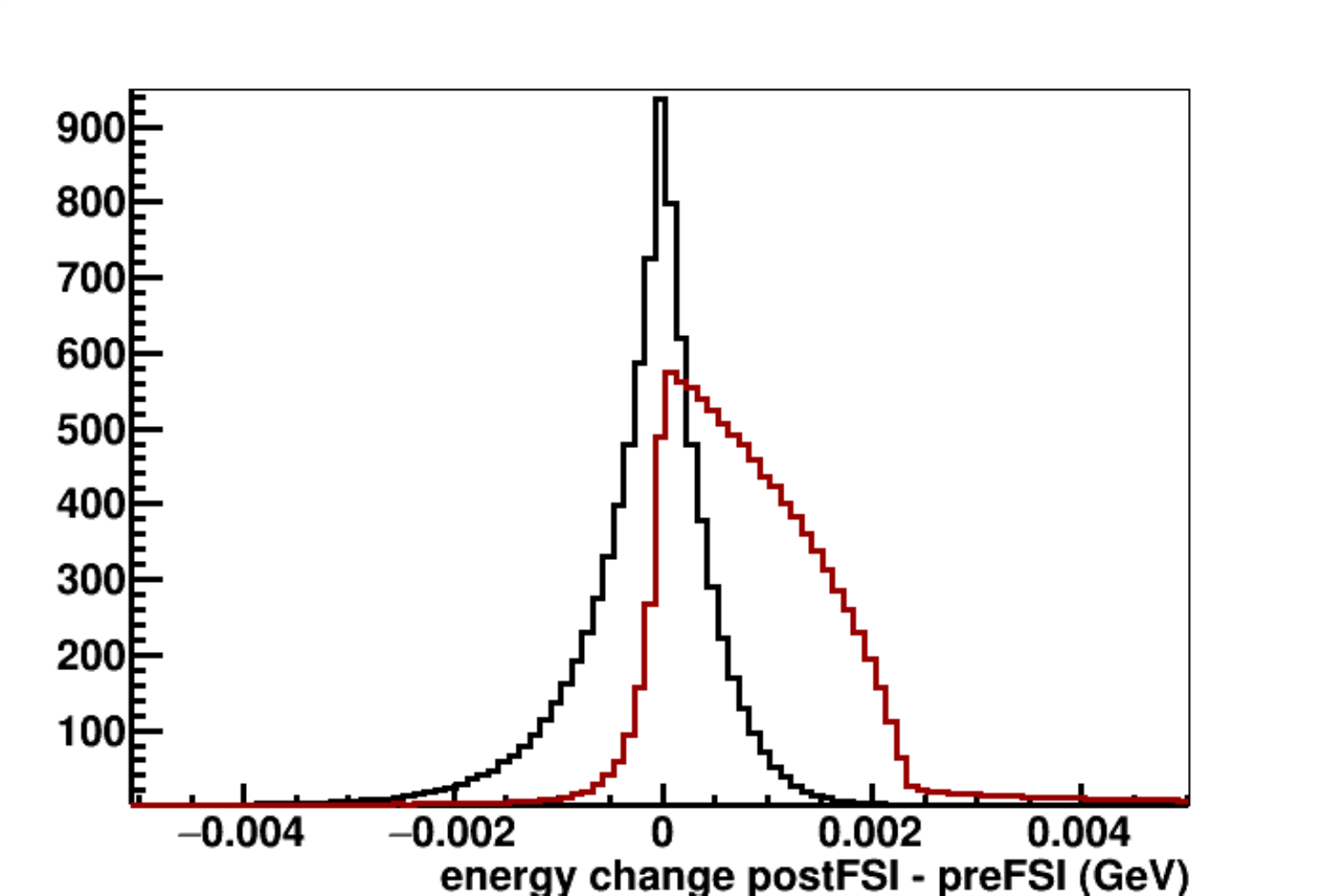}
       \includegraphics[width=0.3\textwidth]{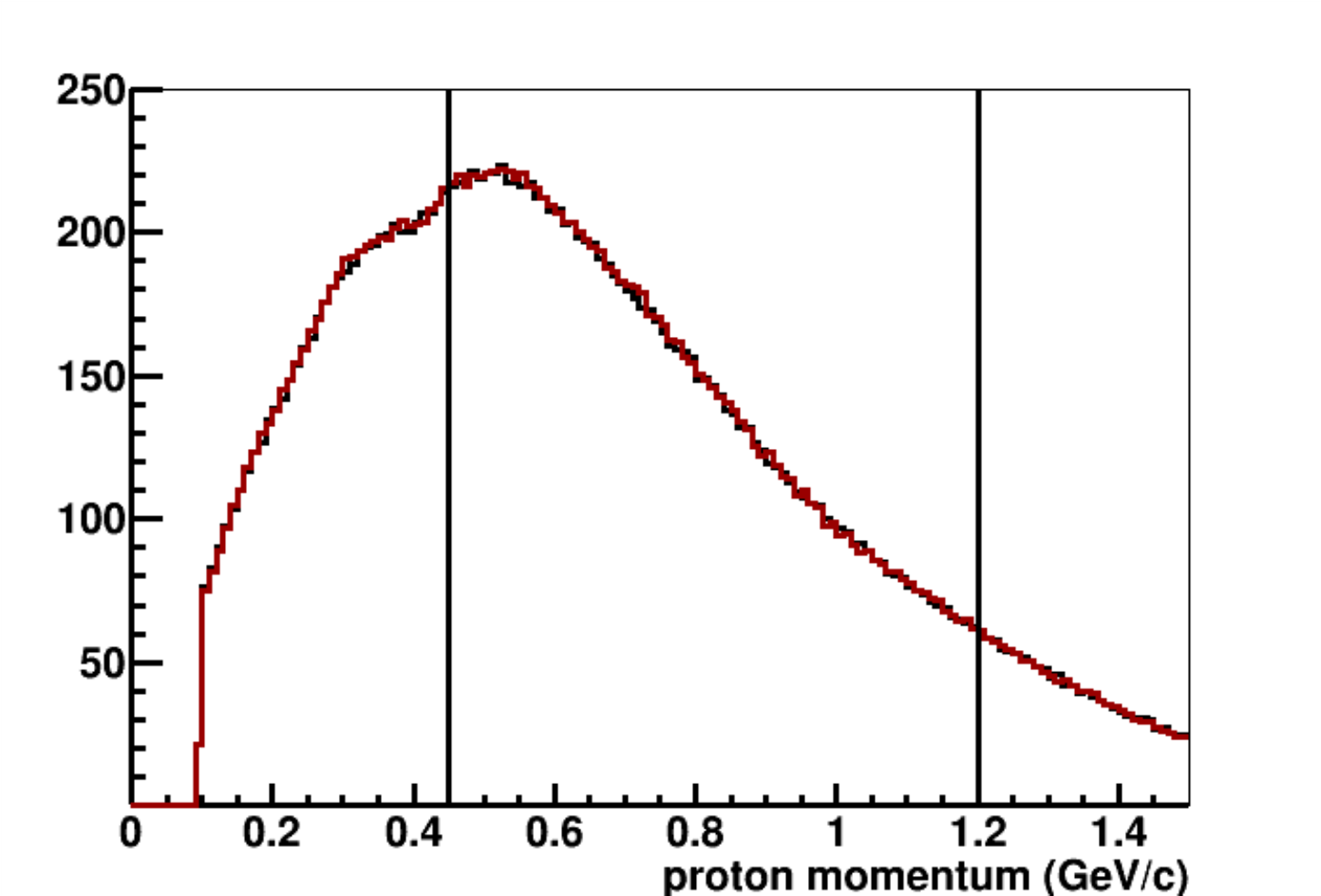}
       \includegraphics[width=0.3\textwidth]{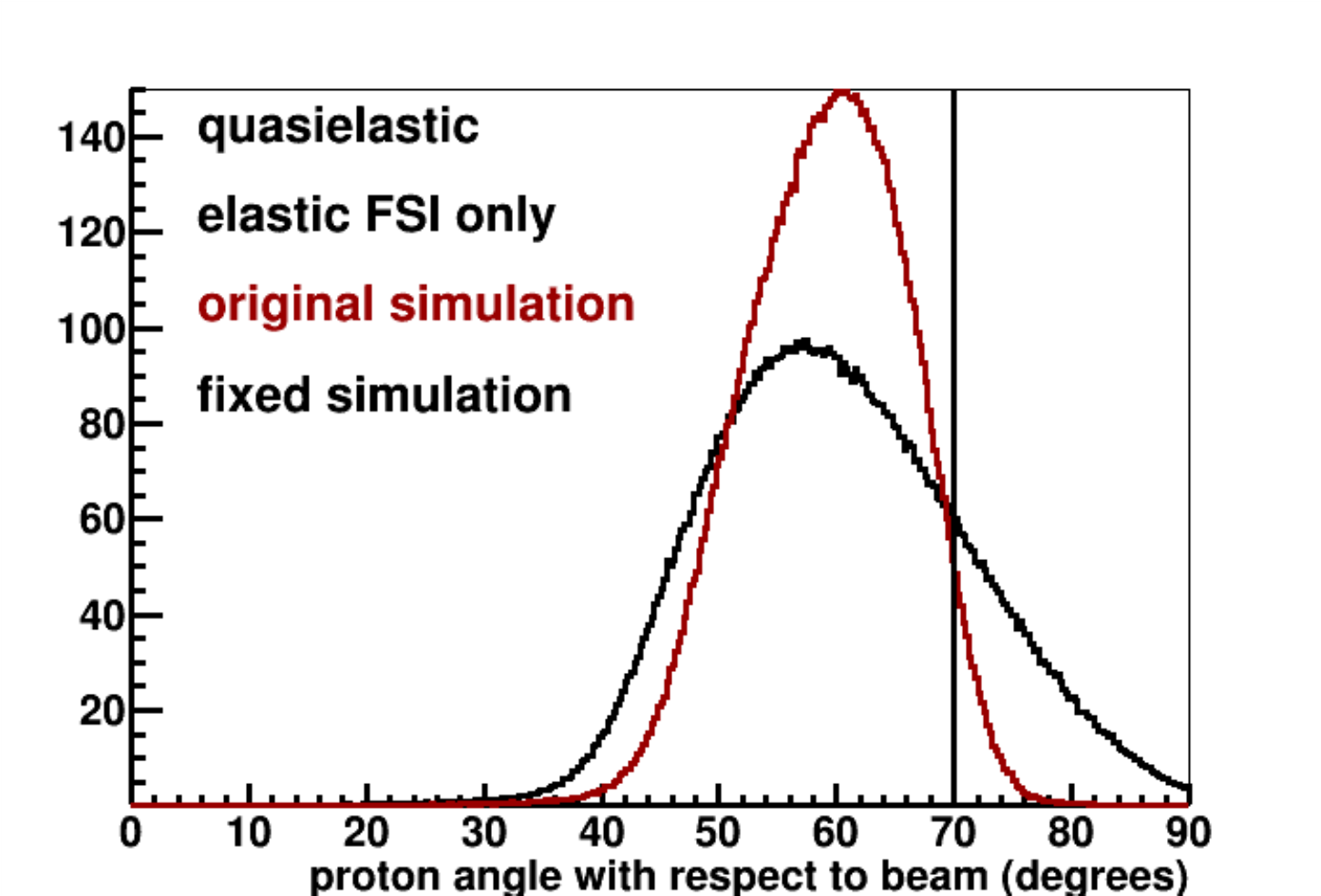}
       \caption{Comparison of the effect of fixing the \genie\ code and comparing key QE proton distributions. Left: Proton acceleration showing the old code produced less than 2 MeV shift. Middle: the small energy shift has negligible effect on the momentum distribution. Right: a major distortion of the angle distribution is what affects the \stki\ analyses; the correct angle distribution is similar to protons (not shown) which had no FSI. Before the fix, this and other distributions based on proton angle such as Fig.~\ref{fig:fsi} are too narrowly peaked. ~\cite{Harewood} }
       \label{fig:lauren}
   \end{figure*}
  
   The prediction from \mnvgenie{1}\ in $\dptx$ has three distinct regions shown in Fig.~\ref{fig:fsi}: a non-CCQE tail beyond $|\dptx|\gtrsim0.2\gevc$, a no FSI CCQE dominated region in $0.2\gtrsim|\dptx|\gtrsim0.1\gevc$, which reflects the Fermi momentum, and an elastic FSI peak at $|\dptx|\lesssim 0.1 \gevc$. The \genie\ elastic FSI is sharply peaked and much narrower than the underlying Fermi gas distribution. Since the protons in the elastic FSI peak follow the no FSI distribution before the FSI simulation, we expect the width of the elastic FSI distribution to be at least as large as that of the no FSI distribution. 
   
   Hints of the unphysical nature of the angular distribution already appeared in the original \stki\ analysis reported in Ref~\cite{Lu:2018stk}. \minerva\ uses the default \genie~ configuration of version 2.12 which uses the ``hA'' model for FSI.  In this model, every nucleon experiences exactly one of the following fates: 1) no FSI, 2) charge exchange with single nucleon knockout, 3) elastic hadron+nucleus scattering, 4) inelastic single nucleon knockout, 5) multi-nucleon knockout (including pion absorption) and 8) pion production.  An advantage of this model is that a reweighting technique can be used to modify the relative mix of fates without fully regenerating the Monte Carlo samples. This is convenient for studying FSI systematic effects with an analysis, similar to the existing FSI uncertainties available with the \genie\ hA model.\\
   
   The routine used to calculate all FSI reactions involving a two body scatter contains (in \genie\ versions 2.6 to version 3.0.6) a mistake that affects hA fates ``2'' and ``4'' (nucleon knockout, with and without charge exchange) and fate ``3'' (elastic hadron nucleus scattering) for both protons and pions.  Fate ``3'', combined with quasielastic events and \stki\ variables, create the largest in observable distributions\cite{Harewood}. \\
  
   The primary effect is on the angular distribution of the scattered hadrons.  In the QE case the original code causes too few of the most highly-transverse protons, which have low efficiency to be tracked in the \minerva\ planar design. It also produces a population, especially of QE events, with a very narrow angle distribution, and in quantities derived from those angular distributions, like many of the \stki\ observables.  The angular distribution relative to the lepton and other hadrons are separately affected.   This combination affects the predicted distributions presented in Ref~\cite{Lu:2018stk} in multiple ways.  In addition, the resulting hadrons pick up an acceleration of up to $2\mev$.  This is smaller than most hadronic energy uncertainties and has negligible role in selection or calorimetry. Instead it appears as an unphysical population in all \stki\ populations in Ref~\cite{Lu:2018stk}. The largest effects of the distorted input model, after performing the iterative unfolding procedure, are on the acceleration angle and the coplanarity angle.
  
   The study reported in Ref~\cite{Harewood} suggests reweighting up the no FSI fate and removing the elastic fate contributions will sufficiently mimic the proton distributions in a fixed code without having to regenerate all MC. Figure~\ref{fig:lauren} shows the effect of fixing the \genie\ code and comparing key QE proton distributions.
   The reason is that the intended elastic scattering angle for protons and neutrons is always small: 90\% would be less than $8^\circ$.
   For this analysis, the weights are only applied to \genie\ quasielastic events; the distortion of angles for non-quasielastic events with multiple hadrons has a small effect on these distributions.

   \begin{figure*}[tbp]
    \includegraphics[width=.45\textwidth]{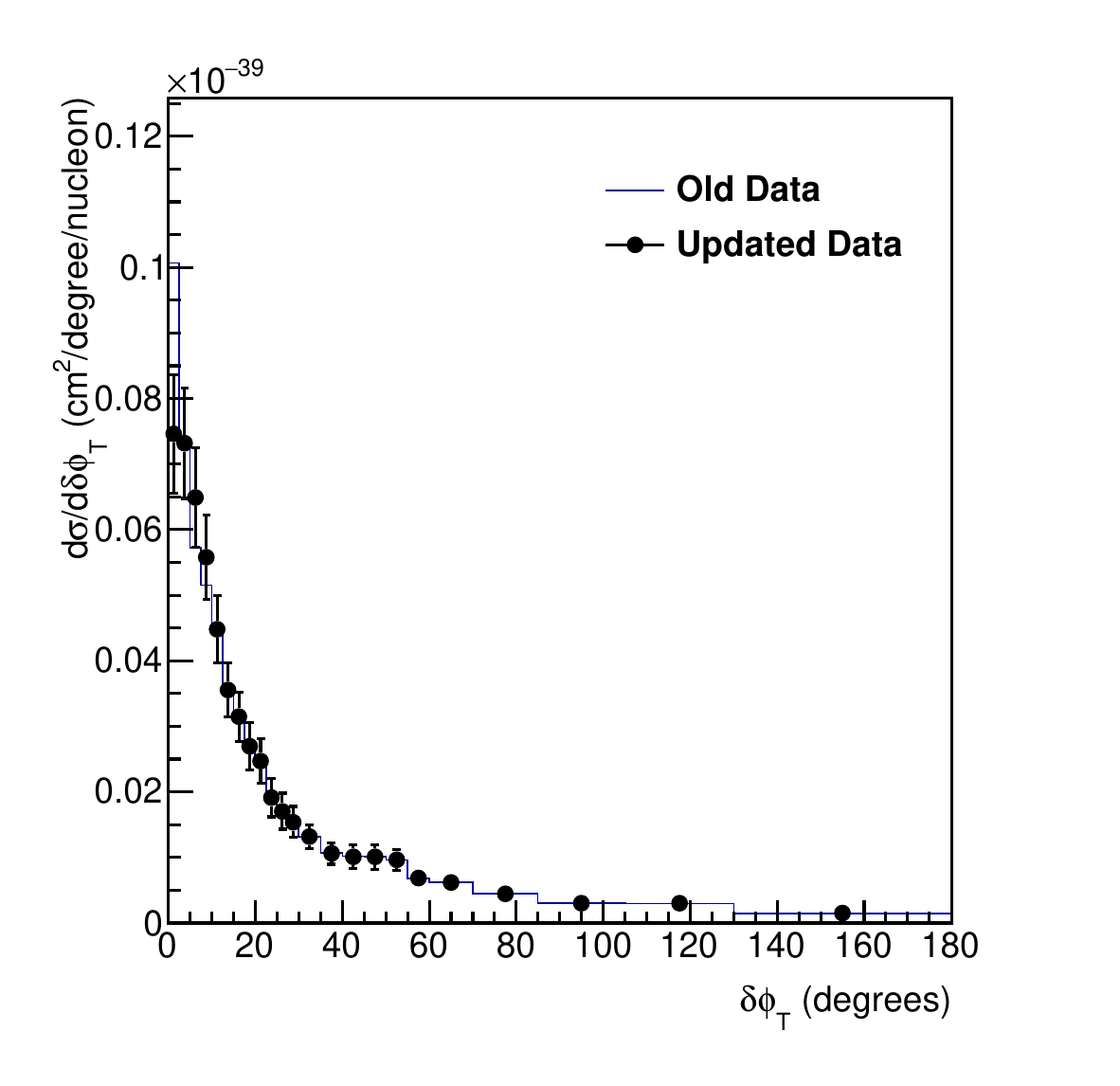}
    \includegraphics[width=.45\textwidth]{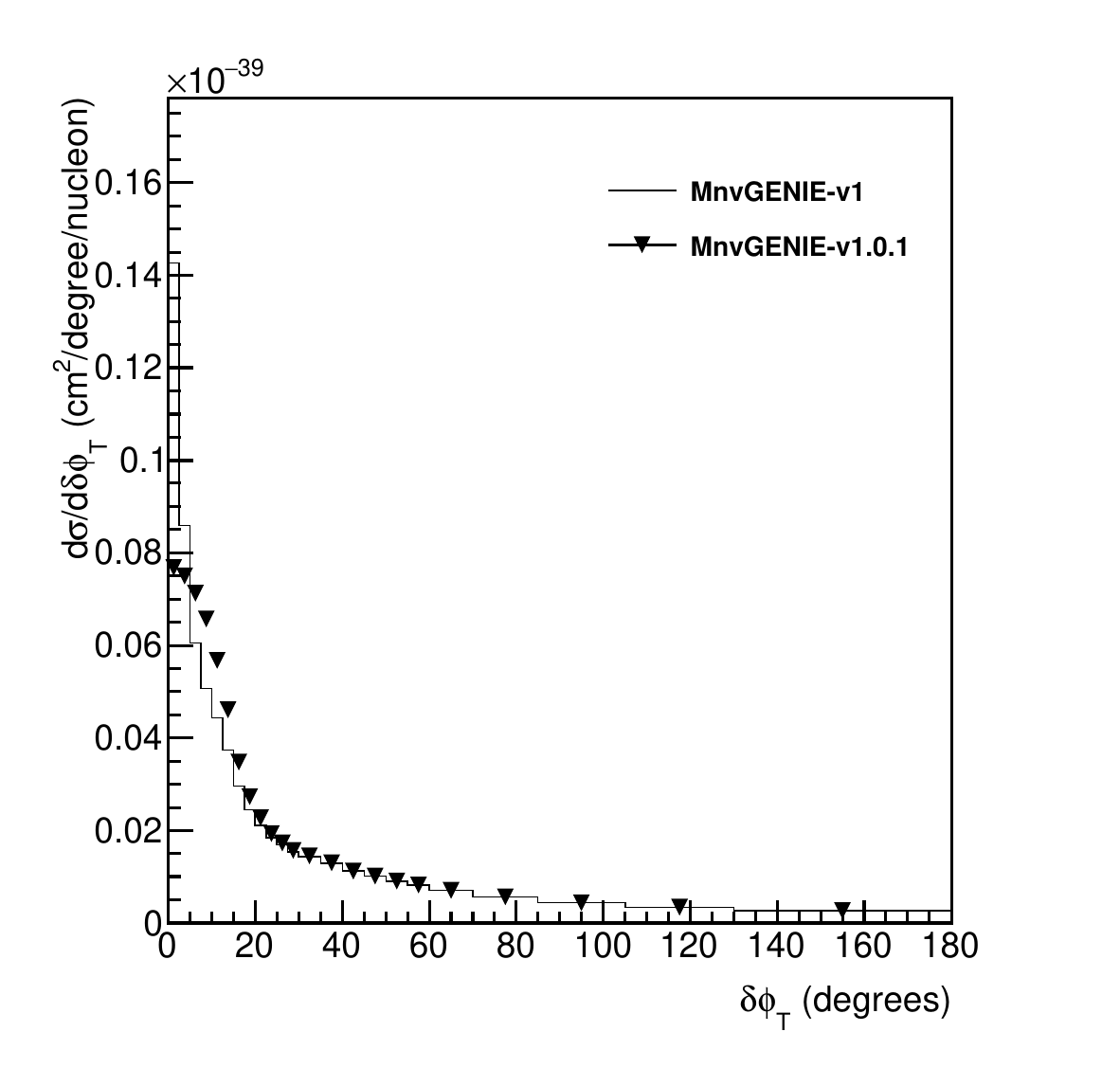}\\
    \includegraphics[width=.45\textwidth]{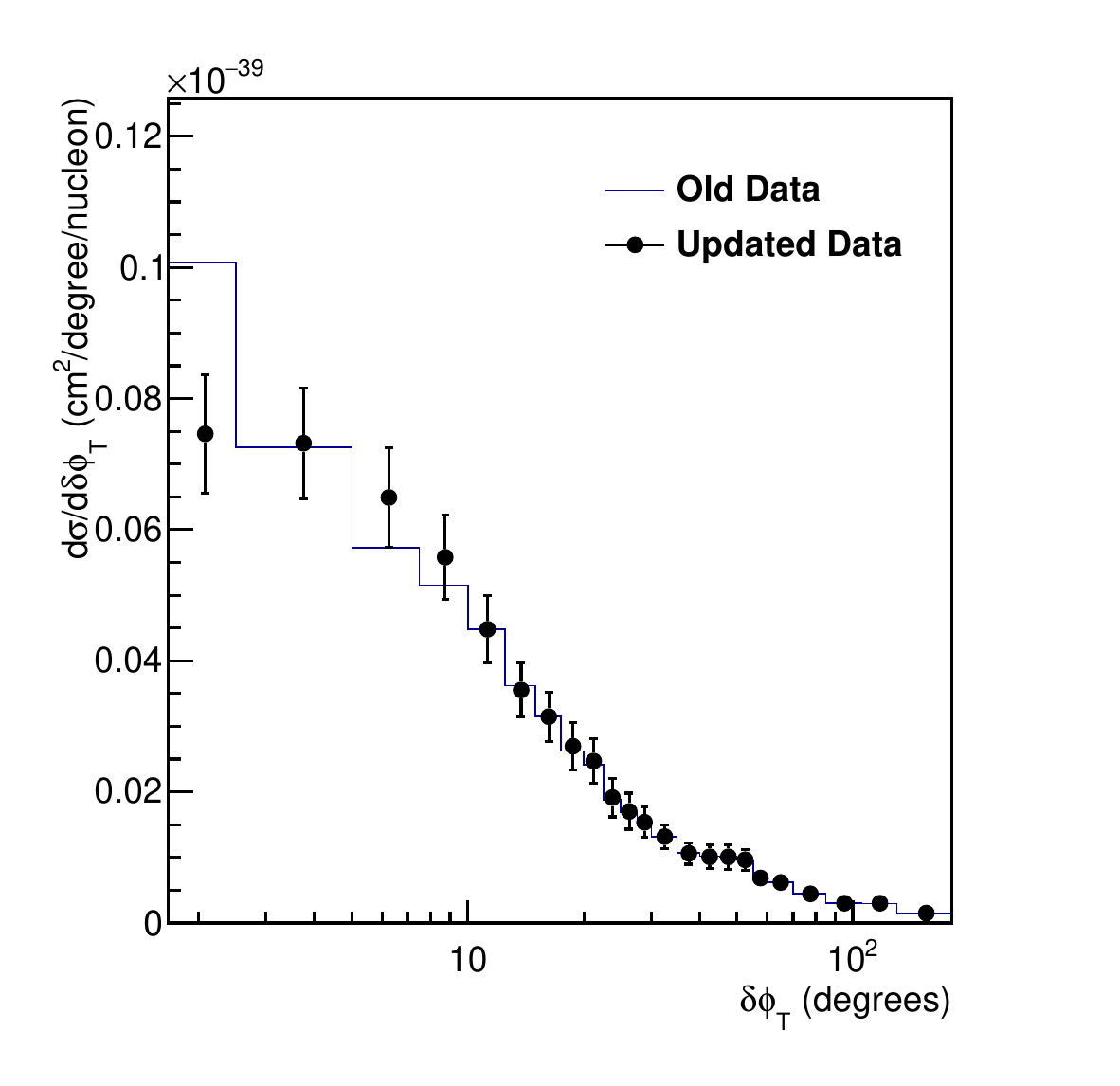}
    \includegraphics[width=.45\textwidth]{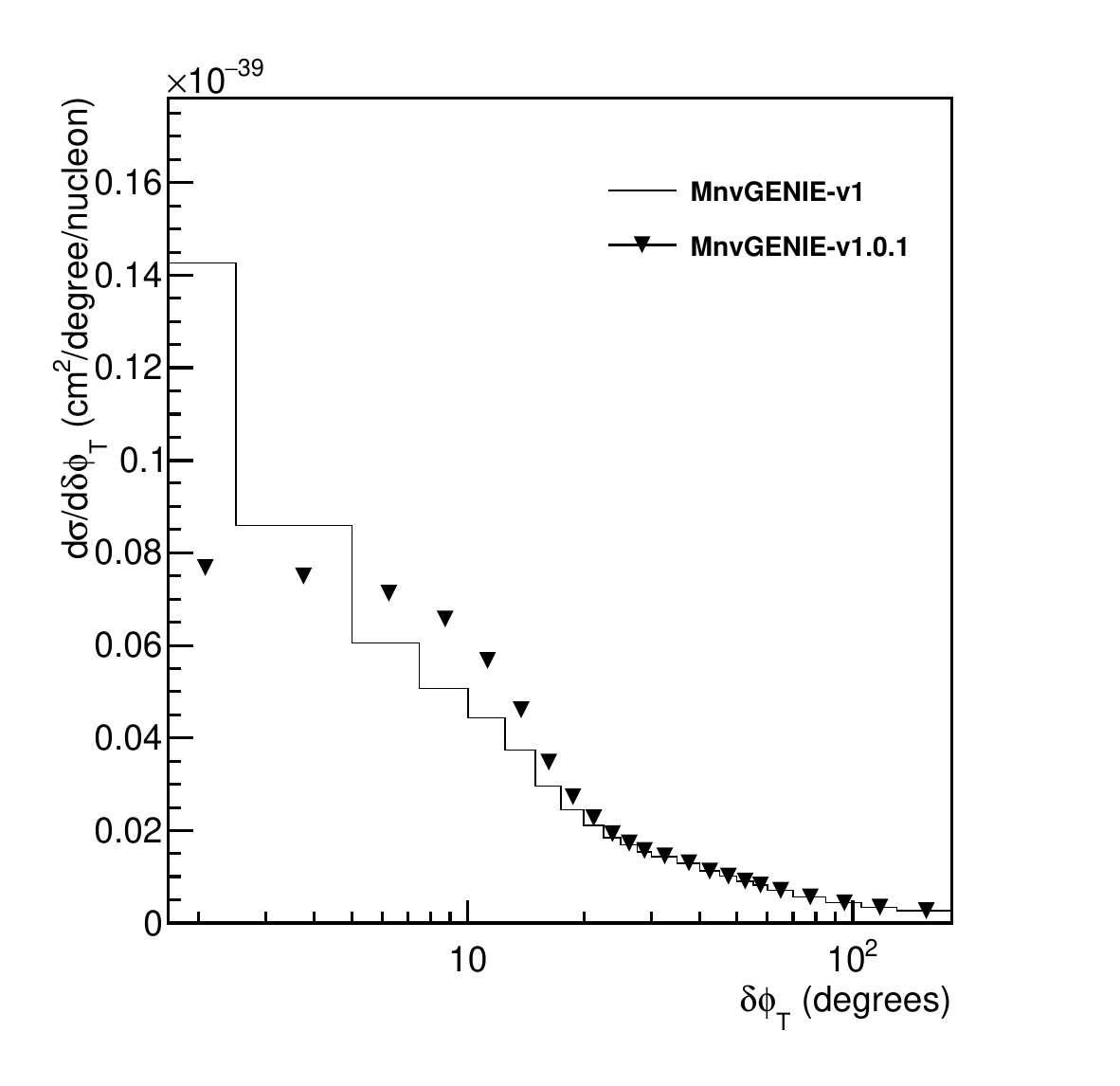}\\
    \caption{$\dphit$ is the angular projection of proton transverse momentum. The original FSI code produces a peak at $0 (rad)$ indicating protons were being produced in the reaction plane more often than Fermi-motion would give.  The first cross section data point is the only one that shifted by more than $2\sigma$.}
    \label{fig:xpphit}
  \end{figure*}  
  
  \begin{figure*}[tbp]
    \includegraphics[width=.45\textwidth]{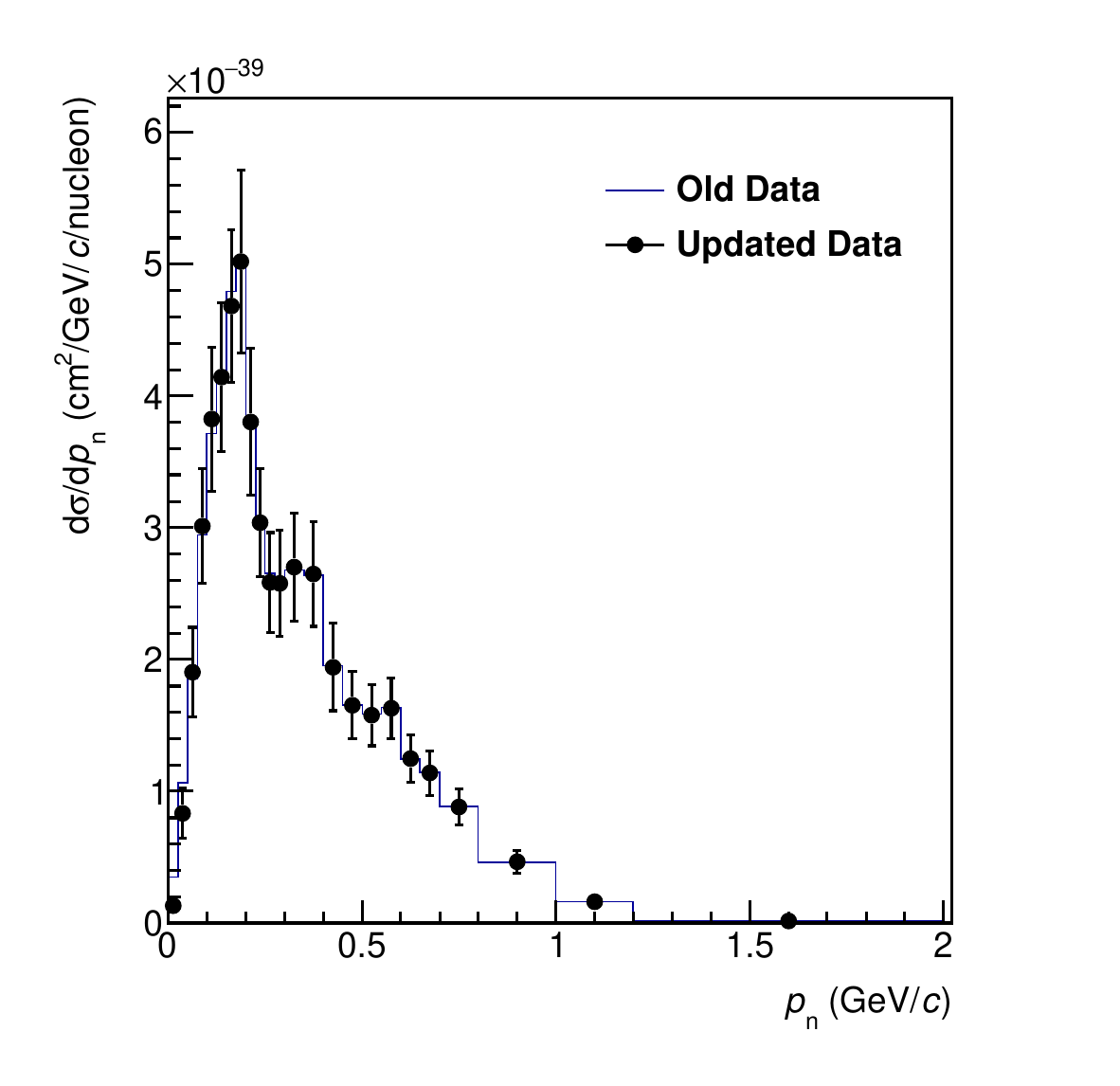}
    \includegraphics[width=.45\textwidth]{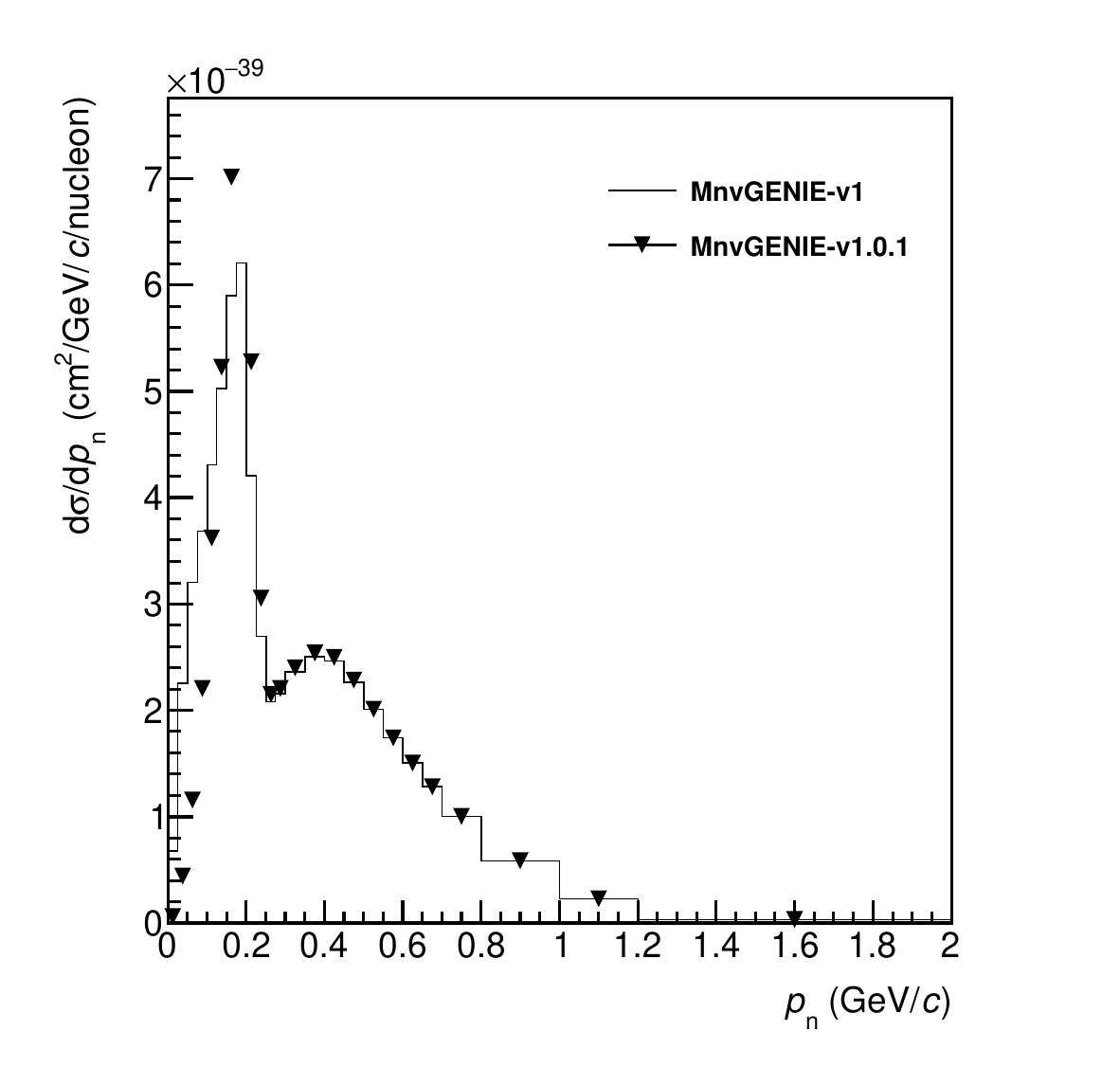}\\
    \includegraphics[width=.45\textwidth]{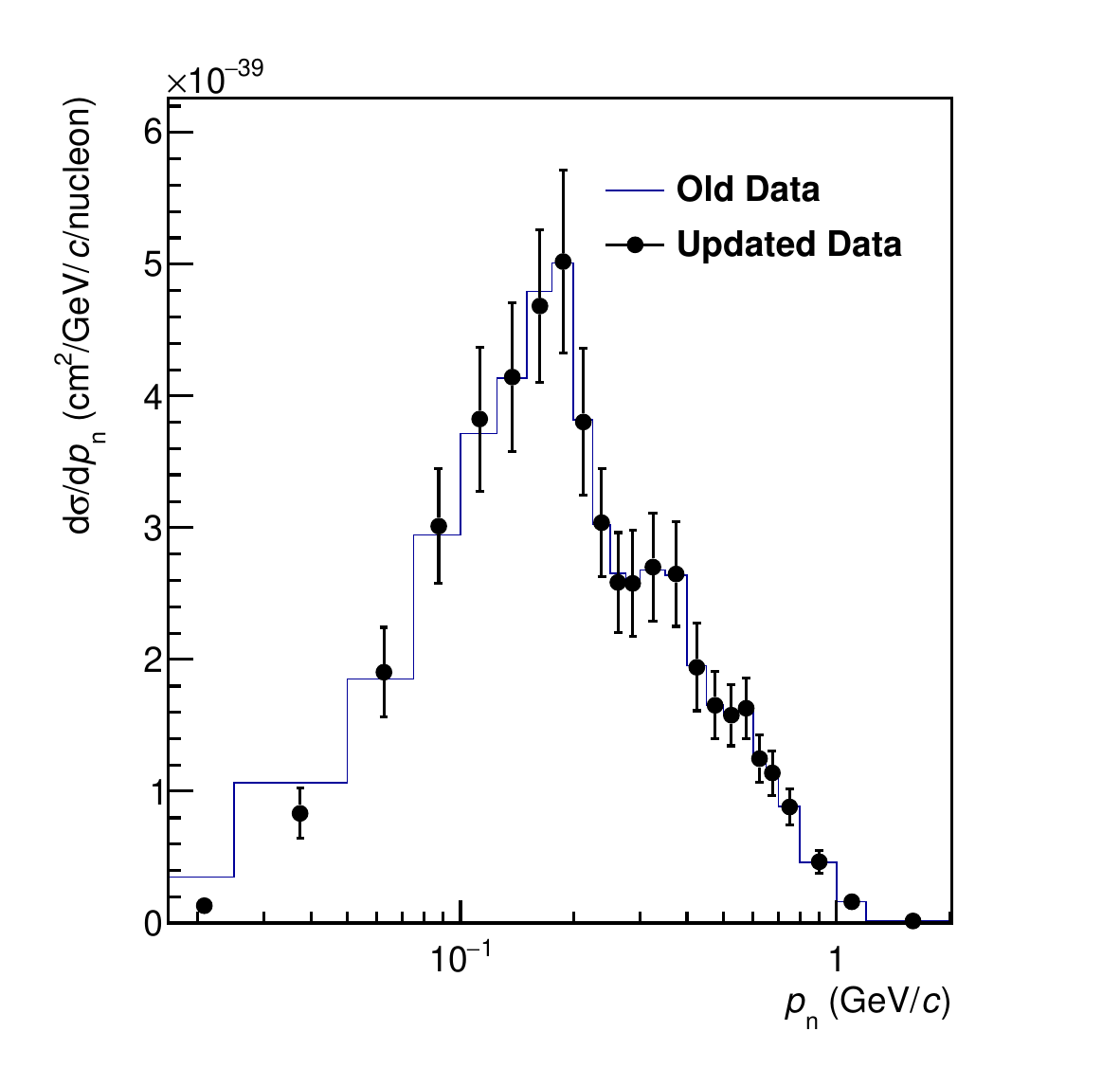}
    \includegraphics[width=.45\textwidth]{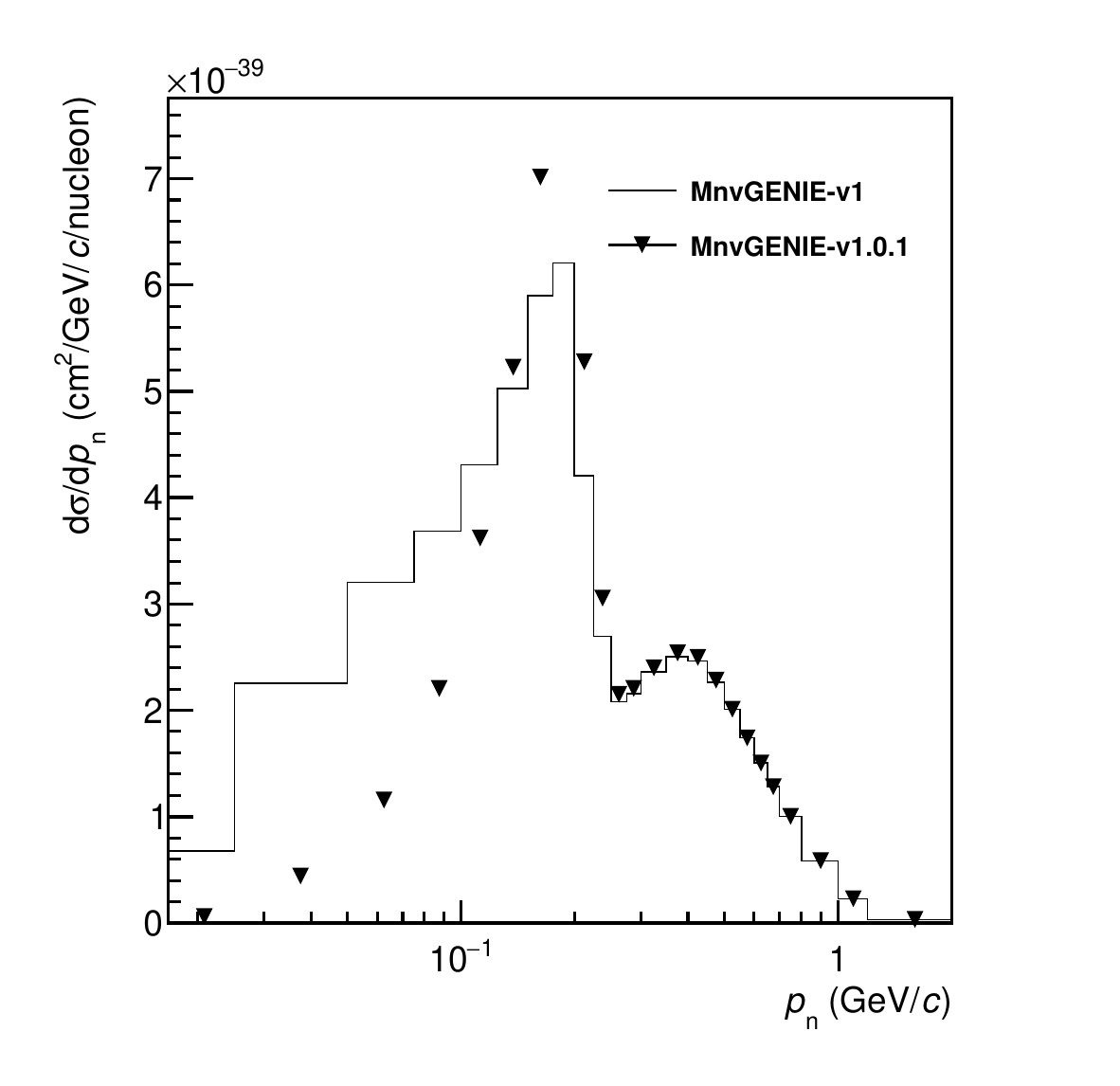}\\
    \caption{ The inferred initial neutron momentum, $\pn$, extracted before and after the elastic FSI reweight. Only the first bin differ more than $2\sigma$. }
    \label{fig:xppn}
  \end{figure*}

   Figure~\ref{fig:xpphit} and ~\ref{fig:xppn} shows $\dphit$ and $\pn$ respectively. The left plot shows the comparison between data extracted using the \mnvgenie{1} (Old Data) and the \mnvgenie{1.0.1} (Updated Data). The right plot compares the two MC models.
   All model distributions are modified significantly, but the extracted cross section shifts are only significant in the first bins of $\dphit$ and $\pn$.

   Table~\ref{tab:chi2vars} compares the old and updated data to the \mnvgenie{1.0.1}\ model. The two data extractions are consistent within 1 unit of reduced $\chi^2$. The two data for $\dphit$ and $\pn$ differ significantly only in the first bin, but the effect on $\chi^2$ is small.
   \begin{table}[bp]
      \caption{$\chi^2$ of the old and updated data compared to \mnvgenie{1.0.1}}
      \label{tab:chi2vars}
       \begin{tabular}{|c|c c|c|c|}\hline
        Variables  & Old Data & Updated Data & DOF & $\Delta$ in Reduced $\chi^2$ \\\hline
        $\pn$       &103        &96.8       &26     &0.24   \\
        $\dalphat$  &25.6       &26.4       &13     &-0.062  \\
        $\dphit$    &100        &77.1       &24     &0.95   \\
        $\dpt$      &48.1       &30.0       &26     &0.70   \\\hline
       \end{tabular}
   \end{table}

  The Supplemental Material to this paper contains an update to the \stki\ results presented in Ref~\cite{Lu:2018stk} based on this modification.
   
  Citation of the new cross sections should include this paper and the original paper~\cite{Lu:2018stk} describing the full method.

\end{document}